\documentclass[aps, reprint, superscriptaddress, nofootinbib, showpacs, floatfix]{revtex4-1}
\pdfoutput=1

\usepackage{graphicx}  % needed for figures
\usepackage{dcolumn}   % needed for some tables
\usepackage{bm}        % for math
\usepackage{amssymb, amsmath}
\usepackage{amssymb,amsmath,mathtools,mathrsfs}
\usepackage{slashed}   % for math
\usepackage{color}
\usepackage[bookmarks]{hyperref}
\usepackage{natbib}
\setcitestyle{numbers,compress,square} % i thought this would not sort the citations, but it does...

\usepackage{tensor} % for dealing with the position of indexes in tensors

\usepackage[usenames,dvipsnames,svgnames,table]{xcolor}

\newcommand{\be}{\begin{equation}}
\newcommand{\ee}{\end{equation}}

\newcommand{\mpl}{M_{\rm Pl}}

\newcommand\alphaB{\alpha_{\text{B}}}
\newcommand\alphaM{\alpha_{\text{M}}}
\newcommand\alphaK{\alpha_{\text{K}}}
\newcommand\alphaT{\alpha_{\text{T}}}
\newcommand\alphaH{\alpha_{\text{H}}}
\newcommand\alphaV{\alpha_{\text{V}}}
\def\d{\delta}

\newcommand{\eqn}[1]{eq.~(\ref{#1})}
\newcommand{\eqns}[2]{eqs.~(\ref{#1}-\ref{#2})}
\newcommand{\appref}[1]{App.~\ref{#1}}
\newcommand{\secref}[1]{Sec.~\ref{#1}}
\newcommand{\half}{\frac{1}{2}}

  \newcommand{\figref}[1]{Fig.~\ref{#1}}

\newcommand{\bun}{\beta_1}
\newcommand{\bdeux}{\beta_2}
\newcommand{\btrois}{\beta_3}

\newcommand{\rhom}{\rho_{\rm m}}
\newcommand{\barrhom}{\bar \rho_{\rm m}}
\newcommand{\kvec}{\vec{k}}
\newcommand{\xvec}{\vec{x}}
\newcommand{\ttt}{ t_1}
\newcommand{\qvec}{\vec{q}}

\begin{document}

\widetext
%\leftline{Saclay-t19/xxx}
%\vspace*{-1cm}
\begin{flushright}
%{\small {\cred Saclay-t19/013}}\\
%\vspace*{2mm} \today
\end{flushright}

\title{{Consistency relations for large-scale structure  in modified gravity \\
and the matter bispectrum }}
\author{Marco Crisostomi}
\affiliation{Institut de physique th\' eorique, Universit\'e  Paris Saclay 
CEA, CNRS, 91191 Gif-sur-Yvette, France}
\affiliation{AIM, CEA, CNRS, Univ.\ Paris-Saclay, Univ.\ Paris Diderot,
Sorbonne Paris Cit\'e, F-91191 Gif-sur-Yvette, France}
\affiliation{Laboratoire de Physique Th\'eorique, CNRS, Univ.\ Paris-Sud, 
Universit\'e Paris-Saclay, 91405 Orsay, France}
\author{Matthew Lewandowski}
\affiliation{Institut de physique th\' eorique, Universit\'e  Paris Saclay 
CEA, CNRS, 91191 Gif-sur-Yvette, France}
\author{Filippo Vernizzi}
\affiliation{Institut de physique th\' eorique, Universit\'e  Paris Saclay 
CEA, CNRS, 91191 Gif-sur-Yvette, France}
    \date{\today}
%\date{\today}

\begin{abstract}
\noindent

We study perturbation theory for large-scale structure in the most general scalar-tensor theories propagating a single scalar degree of freedom, which include Horndeski theories and beyond. We model the parameter space using the effective field theory of dark energy. For Horndeski theories, the gravitational field and fluid equations are invariant under a combination of time-dependent transformations of the coordinates and fields. This symmetry { allows one to construct a physical adiabatic mode which} fixes the perturbation-theory kernels in the squeezed limit and ensures that the well-known consistency relations for large-scale structure, originally derived in general relativity, hold in modified gravity as well. For theories beyond Horndeski, instead, { one generally cannot construct such an adiabatic mode}.  { Because of this}, the perturbation-theory kernels are modified in the squeezed limit and the consistency relations for large-scale structure do not hold. We show, however, that the modification of the squeezed limit depends only on the linear theory. We investigate the observational consequences of this violation by computing the matter bispectrum. In the squeezed limit, the largest effect  is expected when considering the cross-correlation between different tracers. Moreover, the individual contributions to the 1-loop matter power spectrum do not cancel in the infrared limit of the momentum integral, modifying the power spectrum on non-linear scales.  

\end{abstract}

%\pacs{yyy}

\maketitle

%%%%%%%%%%%%%%%%%%%%
%
%
%
\section{Introduction}

Scalar-tensor theories are used as benchmarks to model deviations from  general relativity (GR) in the cosmological context. To avoid instabilities, one usually focuses on the class of theories that propagate a single scalar degree of freedom (and thus  they are free from unstable Ostrogradsky modes \cite{Woodard:2015zca}). Such a class contains Horndeski theories \cite{Horndeski:1974wa,Deffayet:2011gz}, i.e.~scalar-tensor theories with equations of motion that are at most of second-order in the metric and the scalar field. But this class can be extended further by considering higher-order theories that are degenerate \cite{Langlois:2015cwa,Crisostomi:2016czh,BenAchour:2016fzp}, also known as Degenerate Higher-Order Scalar-Tensor (DHOST) theories. Theories beyond Horndeski such as Gleyzes-Langlois-Piazza-Vernizzi (GLPV) theories \cite{Gleyzes:2014dya} belong to this latter class (see also \cite{Zumalacarregui:2013pma} for examples of theories beyond Horndeski).\footnote{Several astrophysical constraints have been recently derived, which put all these theories under pressure. In particular, their parameter space  is constrained \cite{Creminelli:2017sry,Sakstein:2017xjx,Ezquiaga:2017ekz,Baker:2017hug} by
the measurement of the gravitational wave speed \cite{TheLIGOScientific:2017qsa}. The breaking of the  Vainshtein screening inside astrophysical sources for theories beyond Horndeski \cite{Kobayashi:2014ida} allow to further bound the modified gravity parameters \cite{Crisostomi:2017lbg,Langlois:2017dyl,Dima:2017pwp}  (see also \cite{Saltas:2019ius} for a recent improvement of these bounds). Another bound can be established by supressing the decay of gravitational waves predicted by some of these theories \cite{Creminelli:2018xsv,Creminelli:2019nok}. The Vainshtein mechanism for the theories that evade these constraints is discussed in \cite{Hirano:2019scf,Crisostomi:2019yfo}. }

One of the main goals of current and forthcoming cosmological surveys of large-scale structure (LSS)  is to constrain these theories. Beside modifying the linear evolution of perturbations, deviations from GR can also affect higher-order statistics of the cosmic fields and  the formation of structures  in the non-linear regime.   Investigating this regime can be crucial to disentangling the  effects of different theories that are degenerate on linear scales. 

When studying higher-order statistics, it is useful to establish robust relations between correlation functions. The most compelling examples are the consistency relations for LSS in $\Lambda$CDM \cite{Peloso:2013zw,Kehagias:2013yd,Creminelli:2013mca}, which relate an $n$-point function of the density contrast to an $(n+ 1)$-point function in the limit in which one of the
$(n + 1)$ momenta becomes much smaller than the others  (see \cite{Creminelli:2013poa} for an extension of the consistency relations to multiple soft limits and redshift space; see also \cite{Rizzo:2016akm,Rizzo:2017zow} for an example of consistency relations in the late-time Universe involving also the velocity and \cite{Esposito:2019jkb} for a verification of the consistency relations in $N$-body simulations). These relations hold non-perturbatively in the short-scale physics because they 
follow from symmetries of the fluid and gravitational equations. In particular, they can be derived by erasing the effect of a long
mode on the short ones with a suitable combination of coordinate and field transformations \cite{Maldacena:2002vr,Creminelli:2004yq}  based on the equivalence principle \cite{Creminelli:2013mca}. This means that the consistency relations do not hold when the equivalence principle is violated \cite{Creminelli:2013nua}.

A natural question to ask is whether these relations also hold  in modified gravity. Previous calculations of higher-order correlators focused on the 3-point function. In particular, it was found that in Horndeski theories 
the monopole and the quadrupole of the perturbation theory kernel $F_2 (\vec k_1 , \vec k_2)$\footnote{This is explicitly defined in eq.~\eqref{delta2sol2}; as explained below this equation, the kernel $F_2$ can  be organized in terms of {monopole, dipole, and quadrupole terms,} based on their dependence on $ \vec k_1 \cdot \vec k_2$.} get modified (see e.g.~\cite{Bartolo:2013ws, Takushima:2013foa, Bellini:2015oua, Bellini:2015wfa,Burrage:2019afs} for a study of the $F_2$ kernel and the bispectrum in modified gravity; for nonlinear corrections to the bispectrum, see e.g.~\cite{Bose:2018zpk, Bose:2019wuz}) but the dipole remains the same as in $\Lambda$CDM. Indeed, in $\Lambda$CDM the dipole is protected by the symmetry transformations of the fluid and gravitational equations dictated by the {equivalence} principle, the same symmetry transformations at the origin of the consistency relations. This suggests that the symmetry transformations valid for $\Lambda$CDM can be extended to Horndeski theories and that the
consistency relations hold also there. 
On the other hand, it was recently found that in GLPV theories the dipole of $F_2$ gets modified \cite{Hirano:2018uar}, which suggests   that the consistency relations of $\Lambda$CDM  do not hold for theories beyond Horndeski.

In this paper we clarify these statements and show when and why the consistency relations for LSS hold. In the next section we study the gravitational equations and their symmetries. To describe the scalar-tensor theories discussed above, we adopt the Effective Field Theory approach, which  conveniently reduces the number of free time-dependent functions of the parameter space. In Sec.~\ref{fluideqssec}  we extend this discussion to the fluid equations and in Sec.~\ref{symmetrysec} we examine the validity of  the consistency relations, based on the symmetry transformations of the equations established in the previous sections.  In Sec.~\ref{bispectrumsec} we compute the matter bispectrum and the bispectrum involving a different tracer, i.e.~the lensing potential. Other observational consequences of our results are discussed in Sec.~\ref{sec:observations} while Sec.~\ref{sec:conclusion} is devoted to the conclusions. We report the coefficients of the equations used in the text in App.~\ref{sec:coeff}, and \appref{sec:Green} contains the definition of the Green's function and some manipulations useful in the text.

We direct readers primarily interested in the observational consequences of our work to \secref{eomsec} for a description of our model, and then to \secref{bispectrumsec} and \secref{sec:observations} for our main results.

%%%%%%%%%%%%%%%%%%%%%
%
%
%
%
\section{Gravitational Sector} \label{scalarfieldeqssec}

%%%%%%%%%%%%%%
%
\subsection{Action and field equations} \label{eomsec}

We start with the non-linear action that describes DHOST theories in the EFT of {dark energy}  \cite{Dima:2017pwp,Crisostomi:2019yfo} (see \cite{Langlois:2017mxy} for a study of linear perturbations in DHOST theories). { The covariant DHOST action, together with the map to the EFT action, can be found in \appref{dhostactionmap}.}
Using the ADM metric decomposition with  line element  $ds^2 = - N^2 dt^2 + h_{ij} (dx^i + N^i dt) (dx^j + N^j dt)$, and choosing the time as to coincide with the uniform scalar-field hypersurfaces, this reads
\be
\begin{split}
\label{EFTaction}
&S_{\rm EFT} = \int  d^4 x \sqrt{h}  \frac{M^2}{2} \big[ -(1+\delta N)\delta {\cal K}_2 +c_{\rm T}^2 {}^{(3)}\!R \\
& + H^2 \alphaK \delta N^2 + 4 H \alphaB \delta K \delta N   +(1+\alphaH) {}^{(3)}\!R \delta N \\
 &
+4 \beta_1 \delta K V + \beta_2 V^2 + {\beta_3} a_i a^i
+ \alphaV \delta N \delta {\cal K}_2 \big]\;,
\end{split}
\ee
where we have written only the operators with the highest number of spatial derivatives, which are relevant {in the quasi-static limit.}
Here $H\equiv \dot a/a$ (a dot denotes the time derivative), $\delta N \equiv N-1$, $\d K_i^j \equiv K_i^j- H \delta_i^j$  is the perturbation of the extrinsic curvature of the time hypersurfaces, $\delta K$ its trace, and ${}^{(3)}\!R$ is the 3D Ricci scalar of these hypersurfaces. Moreover, $\delta {\cal K}_2 \equiv \delta K^2 - \delta K_{i}^j \delta K^{i}_{j}$, $V \equiv (\dot N - N^i \partial_i N)/N${,} and $a_i \equiv \partial_i N/N$. { Forecasted limits on these parameters with future large-scale structure observations give $\alpha_i \lesssim {\cal O}(0.1)$ (see e.g.~\cite{Frusciante:2019xia} for a recent review).}\footnote{{ The cutoff of these theories can be estimated by computing at what energy perturbative unitarity is violated in the $2 \to 2$ scattering of $\pi$. For the above values of parameters, this gives (see e.g.~\cite{Pirtskhalava:2015zwa}) $\Lambda_\text{cutoff} \sim (10^3\, \text{km})^{-1}$. Note however that analyticity arguments suggest a much lower energy, i.e.~$\Lambda_\text{cutoff} \sim (10^7\, \text{km})^{-1}$ \cite{Bellazzini:2019xts}. Anyway, these correspond to much smaller scales than those considered in this paper, which are roughly near $k_{\rm LSS} \sim (10^{20} \, \text{km} )^{-1}$.}}
For $\alphaH = \beta_1=\beta_2=\beta_3=0$ this action describes Horndeski theories. In this case there are four free time-dependent functions: $\alphaK$, $\alphaB$, $c_{\rm T}^2$ and $\alphaM \equiv d \ln M^2/d \ln a$  \cite{Bellini:2014fua}, where $a$ is the scale factor of the homogenous FRW background $ds^2 = - dt^2 + a^2(t) d \vec x^2$. DHOST theories are described also by $\alphaH$ and $\beta_1$, while the functions $\beta_2$ and $\beta_3$ are given in terms of $\beta_1$ by the degeneracy conditions  \cite{Langlois:2017mxy} $\bdeux=-6\bun^2$,  $ \btrois=-2\bun\left[2(1+\alphaH)+\bun c_{\rm T}^2 \right]$ ,
which we will always impose.

We also assume that matter is minimally coupled to the gravitational metric $g_{\mu \nu}$, and a universal coupling of all species. For simplicity we will focus on non-relativistic matter with vanishing pressure.  

To study cosmological perturbations  we  abandon the unitary gauge by performing a space-time dependent shift in the time $t \rightarrow t + \pi ( t , \vec{x})$, 
and work in the Newtonian gauge,  with metric
\be \label{newtongauge}
ds^2 = - ( 1 + 2 \Phi ) dt^2 + a^2(t) ( 1 - 2 \Psi ) d \vec{x}^2 \ .
\ee
Then, we expand the action \eqn{EFTaction} in terms of the metric and scalar field perturbations $\pi$ and keep only terms with the highest number of derivatives per field, which are those relevant in the quasi-static limit.
The gravitational equations are obtained by varying the action with respect to $\Phi$, $\Psi$ and $\pi$. The detailed procedure can be found in \cite{Crisostomi:2019yfo}.  
Because we are interested in the bispectrum, we need the equations 
up to second order.  In terms of the matter overdensity $\delta \equiv \delta \rho_{\rm m} / \bar \rho_{\rm m}$ (here $\rhom$ is the matter energy density, and $\barrhom$ is its mean cosmological value), these are given by 
\begin{align} \label{phieq1}
\begin{split}
& \frac{a^2 \bar \rho_{\rm m} \delta }{2 M^2} = C_1\partial^2 \pi - \frac{c_8}{4}  \partial^2 \dot \pi + \frac{c_6}{2} \partial^2 \Phi   + \frac{c_4}{4}  \partial^2 \Psi 
\\
& \hspace{.25in} +  \frac{1}{4} \left[  \frac{b_2}{a^2} Q_2 [\pi,\pi ] + \frac{c_8}{a^2} \partial_i \left( \partial_j \pi \partial_i \partial_j \pi \right) \right] \ , 
\end{split}
\end{align}
\begin{align} \label{psieq1}
\begin{split}
&  0 = C_2 \partial^2 \pi - \frac{c_7}{4}  \partial^2 \dot \pi + \frac{c_4}{4}  \partial^2 \Phi  + \frac{c_5}{2}  \partial^2 \Psi \\
& \hspace{.5in} +  \frac{1}{4} \left[  \frac{b_3}{a^2} Q_2 [\pi,\pi ] + \frac{c_7}{a^2} \partial_i \left( \partial_j \pi \partial_i \partial_j \pi \right) \right] \ , 
\end{split}
\end{align}
and 
\begin{align}\label{pieq1}
\begin{split}
 &  0= C_3  \partial^2 \pi +  C_4 \partial^2 \dot \pi + \frac{c_9}{2} \partial^2 \ddot \pi  +  \frac{c_1}{4}  \partial^2 \Phi     + \frac{c_8}{4}  \partial^2 \dot \Phi \\
&+  \frac{c_2}{4}  \partial^2 \Psi + \frac{c_7}{4}  \partial^2 \dot \Psi    +  \frac{1}{4 a^2}  Q_2 [ \pi , b_1\pi+ 2b_2 \Phi + 2b_3 \Psi ] \\
&  - \frac{1}{4 a^2} \partial_i \left[ \partial_i \pi  \, \partial^2 ( c_7 \Psi + c_8 \Phi  + 2 c_9 \dot \pi) \right] \\
&   + \frac{ H c_9 - C_4 }{2 a^2}   \partial^2 \left( \partial \pi \right)^2  - \frac{c_9}{2 a^2 } \partial^2 (  \partial_i \pi \partial_i \dot \pi  )  \ ,
\end{split}
\end{align}
where we have defined 
\be
Q_2 [ \varphi_a , \varphi_b]  \equiv \varepsilon^{ikm} \varepsilon^{jlm} \partial_i \partial_j \varphi_a \partial_k \partial_l \varphi_b \;,
\ee
with 
\be
\varphi_a \equiv \{ \Phi, \Psi, \pi\}  \;,
\ee
and $C_1, \dots, C_4$, 
 $c_1, \ldots, c_9$, and $b_1, b_2, b_3$ are time-dependent coefficients that depend on the parameters of the action, reported in \appref{coefficientsapp}.

%%%%%%%%%%%%%%%%
%
%
\subsection{Perturbative solutions}

In this section we seek a perturbative solution to {eqs.~(\ref{phieq1} -  \ref{pieq1}) } in powers of $\delta$. Thus, we will expand the fields $\varphi_a$ as
\be
\varphi_a = \varphi_a^{(1)} + \varphi_a^{(2)} + \cdots \;,
\ee 
 where each perturbative piece is proportional to the relevant number of powers of  $\delta^{(1)}$, i.e. $\varphi_a^{(n)} \sim [\delta^{(1)}]^n$.  Since we are interested in the bispectrum in this work, we will solve up to second order.

%%%%%%%%%%%%%%%
\subsubsection{Linear solutions}

As discussed in \cite{Crisostomi:2019yfo} (see also \cite{Hirano:2018uar,Crisostomi:2017pjs,Hirano:2019nkz,Hirano:2019scf}), the linear solutions have the following form
\begin{align}
\begin{split} \label{linearsols}
 a^{-2} \partial^2 \varphi_a^{(1)}  = \mu_{\varphi_a} \delta^{(1)} + \nu_{\varphi_a} \dot \delta^{(1)} + \sigma_{\varphi_a} \ddot \delta^{(1)}  \  ,
\end{split}
\end{align}
and we have supplied the expressions for the time dependent $\mu_{\varphi_a}$, $\nu_{\varphi_a}$, and $\sigma_{\varphi_a}$ functions in terms of the parameters in the 
field equations
in \appref{linearsolssec}.  We note, however, that 
\be
\sigma_\pi = 0 \ .
\ee   
Horndeski theories have $\sigma_{\varphi_a} = \nu_{\varphi_a} = 0$, and $\Lambda$CDM has $\mu_\Phi = \mu_\Psi = \bar \rho_{\rm m} / ( 2 \mpl^2)$, where $\mpl$ is the Planck mass.

In \secref{fluidseclin} we will derive the evolution equation for $\delta^{(1)}$ in closed form. In the quasi-static approximation it is scale independent in these theories. In particular, its solution can be written in the form
\be \label{lineardelta}
\delta^{(1)} ( \xvec , t ) = \frac{D_+ ( t )}{D_+ ( t_{\rm in} ) } \delta^{(1)} (\xvec,  t_{\rm in} ) \ , 
\ee
where $D_+$ is the linear growth factor and $t_{\rm in}$ is some early time where we set the initial conditions.  This means that we can write time derivatives of $\delta^{(1)}$ as proportional to $\delta^{(1)}$, i.e.
$ \dot \delta^{(1)} = H f \delta^{(1)} $, 
where 
\be 
\label{fdef}
f \equiv \frac{\dot D_+ }{  H D_+}  \;,
\ee
is the linear growth function.  

{Once one has the linear solution $D_+$ from \secref{fluidseclin}}, this allows us to write the linear solutions \eqn{linearsols} as 
\begin{align}
\begin{split} \label{Ldefs1}
a^{-2} \partial^2 \varphi_a^{(1)} =  L_{\varphi_a} \delta^{(1)} \ , 
\end{split}
\end{align}
where
\begin{align}
\begin{split} \label{Ldefs}
L_{\varphi_a}  = \mu_{\varphi_a} + H f \nu_{\varphi_a} + (H^2 f^2 + H \dot f + \dot H f ) \sigma_{\varphi_a}  \ . 
\end{split}
\end{align}

%%%%%%%%%%%%%%%
%
\subsubsection{Second-order solutions}

To find the second-order solutions, we first solve for 
$\varphi_a^{(2)}$ 
to second order in the potentials.  Then, in the quadratic terms we can use 
the linear solution of the field equations, formally given by \eqn{Ldefs1}, to write the potentials in terms of $\delta^{(1)}$ only.  
After doing this, the solutions for the potentials have the form
\be
\label{secondorderpoisson}
a^{-2} \partial^2  \varphi_a^{(2)} \equiv \mu_{\varphi_a} \delta^{(2)} + \nu_{\varphi_a} \dot \delta^{(2)} + \sigma_{\varphi_a} \ddot \delta^{(2)} + a^{-2} \partial^2 \varphi_a^{(2),{\rm NL}} \ , 
\ee
where the last term on the right-hand side is quadratic in $\delta^{(1)}$ and is given by
\be \label{secondorderform}
a^{-2} \partial^2 \varphi_a^{(2) ,{\rm NL}}= \upsilon_\alpha^{\varphi_a} X_{\alpha_s}  + \upsilon_\gamma^{\varphi_a} X_\gamma   \ .
\ee
Here $\upsilon_\alpha^{\varphi_a} $ and $\upsilon_\gamma^{\varphi_a}$ are time dependent functions given explicitly in \appref{quadraticapp}, and 
\begin{align}
\begin{split} \label{xdefs}
X_{\alpha_s}   & \equiv \big(\delta^{(1)} \big)^{2} + \left(\frac{\partial_i}{\partial^2 } \delta^{(1)} \right) \, \big( \partial_i \delta^{(1)} \big)  \ ,  \\
X_\gamma     & \equiv \big(\delta^{(1)} \big)^{2} - \left( \frac{ \partial_i \partial_j }{\partial^2} \delta^{(1)} \right)^2   \ ,
\end{split}
\end{align} 
are 
two
types of non-linear mixing terms that come from the interactions present in the 
field equations,
\eqns{phieq1}{pieq1}.  Another combination, $X_\beta$,  can also appear but this is 
simply a linear combination of the other two, 
$X_\beta = X_{\alpha_s} -X_\gamma$.  
Note that $\upsilon_\alpha^{\varphi_a}=0$ in Horndeski theories.
In Fourier space, these interactions become the familiar vertices in perturbation theory for LSS.
Using the following notation for the Fourier integrals,
\be
\int^{\kvec}_{\kvec_1 , \cdots , \kvec_n} = \int \frac{d^3 k_1}{(2 \pi)^3}\dots \frac{d^3 k_n}{(2 \pi)^3} ( 2 \pi )^3 \delta_D ( \kvec - \sum_{i=1}^n \kvec_i ) \;, 
\ee
where $\delta_D$ is the Dirac delta function, in Fourier space these read
\be
\begin{split}
\hat X_{\alpha_s}  (\kvec)  & =   \int^{\kvec}_{\kvec_1,\kvec_2} \, \alpha_s ( \kvec_1 , \kvec_2 )  \delta^{(1)}  ( \kvec_1 ) \delta^{(1)} ( \kvec_2)  \;,  \\
\hat X_{\beta}  (\kvec)  & =   \int^{\kvec}_{\kvec_1,\kvec_2} \, \beta ( \kvec_1 , \kvec_2 )  \delta^{(1)} ( \kvec_1) \delta^{(1)} ( \kvec_2 )   \;, \\
\hat X_{\gamma} (\kvec)  & =   \int^{\kvec}_{\kvec_1,\kvec_2} \, \gamma ( \kvec_1 , \kvec_2 )  \delta^{(1)} ( \kvec_1 ) \delta^{(1)} ( \kvec_2 )  \;, \\
\end{split}
\ee
where 
\begin{align} \label{alphakernel}
\alpha_s(\kvec_1 , \kvec_2)  &  \equiv 1 + \frac{ \hat k_1 \cdot \hat k_2}{2} \left(  \frac{k_2}{k_1} + \frac{k_1}{k_2}   \right) \ , \\
\beta(\kvec_1 , \kvec_2)  &  \equiv \frac{ \hat k_1 \cdot \hat k_2}{2} \left(  \frac{k_2}{k_1} + \frac{k_1}{k_2}   \right)  + \left( \hat k_1 \cdot \hat k_2 \right)^2  \ , 
\end{align}
are the usual (symmetrized) perturbation theory kernels \cite{Bernardeau:2001qr} and 
\be
  \gamma ( \kvec_1 , \kvec_2)   \equiv 1 - \left( \hat k_1 \cdot \hat k_2 \right)^2  \;
\ee
is a kernel appearing in modified gravity models (see e.g.~\cite{Cusin:2017mzw,Cusin:2017wjg}).

%%%%%%%%%%%
%
%
%%%%%%%%%%%
%
%
% 
\subsection{Symmetries of the field equations and infrared behavior} \label{leadingirtermssec}
The gravitational field equations, \eqns{phieq1}{pieq1}, 
are invariant under the following coordinate change and shifts of the fields:  
\begin{align}
\begin{split} \label{transf1}
  \tilde x^i &=  x^i +  \xi^i ( t )  \ ,  \quad {\tilde t =  t  \ ,}  \\
\tilde \varphi_a ( \tilde x^j , \tilde t ) & = \varphi_a ( x^j , t) + b^i_{\varphi_a} ( t ) \tilde  x^i   \;, \\
 \tilde \delta ( \tilde x^j ,\tilde  t ) & = \delta ( x^j , t ) \ .
\end{split}
\end{align}
Equivalently,  they are invariant under the  replacements
\begin{align}
\begin{split}  \label{newtransf1}
\partial_i & \rightarrow \partial_i  \;, \qquad \partial_t   \rightarrow \partial_t  - \dot \xi^i ( t )   \partial_i   \;, \\ 
\varphi_a  & \rightarrow \varphi_a + b^i_{\varphi_a} ( t ) x^i \ , \qquad \delta  \rightarrow \delta \;.
\end{split}
\end{align} 

For Horndeski theories (i.e., when $\alphaH = \beta_1 = 0$), this symmetry holds for arbitrary time-dependent functions $\xi^i ( t )$, $b^i_\Phi ( t )$,  $b^i_\Psi ( t )${,} and  $b^i_\pi ( t )$. This is easy to verify. Indeed,  in this case $c_6=c_7=c_8=c_9=C_4=0$ in these equations so that time derivatives are absent and all fields have at least two spatial derivatives.

For DHOST theories (i.e., when $\alphaH \neq 0$ or $\beta_1 \neq 0$), however, the field equations are only invariant 
as long as 
\be \label{bipi}
b^i_\pi ( t ) = - a^2 \dot \xi^i ( t ) \ .
\ee 
Indeed, in this case time derivatives and terms with one spatial derivative on a field are present. 
By the transformation \eqn{transf1} in the form of \eqn{newtransf1}, a time derivative of a field generates a quadratic term involving that field and $\dot \xi^i$.  This term can only be canceled by another quadratic term involving $\partial_i \pi$ if \eqn{bipi} holds.

We will now show that this symmetry of the equations determines the leading {infrared (IR)} behavior,  or squeezed limit, 
of the last term of the second-order field solutions, eq.~\eqref{secondorderpoisson}, i.e.~$\varphi_a^{(2),{\rm NL}}$. This limit is obtained by making an expansion in terms of $q / k$, where $q$ is the wavenumber of a long mode, and $k$ is that of a short mode, i.e.~$q \ll k$.
In this limit, $X_{\alpha_s}$ is enhanced with respect to the other mixing terms, and  eq.~\eqref{secondorderform} gives 
\be \label{phi2nllimit}
a^{-2} \partial^2 \varphi_{a}^{(2),{\rm NL}} \approx  \upsilon^{\varphi_a}_\alpha \frac{\partial_i \delta_L^{(1)}}{\partial^2} \partial_i \delta^{(1)}  + {\cal O}  \left( (q/k)^0\right) \;,
\ee
where, in Fourier space, the leading term on the right-hand side starts at order $(q/k)^{-1}$. 
Here and in the following we will use the  symbol $\approx$ to denote an equality that is valid at leading order in the squeezed limit.
As we will see below, this limit determines the dipole term in the second-order perturbation theory kernel $F_2$. 

One can imagine $\xi^i(t)$ to have a weak $k$-dependence, i.e.~to be a long mode. In this case the transformation \eqn{transf1} captures the leading dependence of that long mode.  
Indeed, this symmetry, together with the condition in \eqn{bipi}, implies that  time derivatives of a field must always come in combination with specific non-linear terms. 
%To see how this works, consider the quadratic term proportional to $\dot \xi^i$ generated by applying \eqn{newtransf1} on a time derivative.
%This term can only be canceled by another term involving $\partial_i \pi$, because of \eqn{bipi}.  Thus, in order for the equations to be invariant under \eqn{newtransf1}, time derivatives of a field must always come in combination with specific non-linear terms. 
% 
 In particular, in \eqn{secondorderpoisson}  the specific non-linear terms generated by transforming $\dot \delta^{(2)}$ and $\ddot \delta^{(2)}$ under \eqn{newtransf1} must be canceled by specific non-linear terms contained in 
  $a^{-2} \partial^2 \varphi_{a}^{(2),{\rm NL}}$. 
Using that under \eqn{newtransf1}
\begin{align}
 \label{deltadottrans}
&  \dot \delta^{(2)}  \rightarrow\dot \delta^{(2)} - a^{-1} v_{\pi,L}^i \partial_i \delta^{(1)} \ , \\
& \ddot \delta^{(2)}  \rightarrow \ddot \delta^{(2)} - a^{-1} \left( 2 v_{\pi,L}^i \partial_i \dot \delta^{(1)} + \dot v_{\pi,L}^i \partial_i \delta^{(1)} - H v_{\pi,L}^i \partial_i \delta^{(1)} \right)  \ , \nonumber
\end{align}
up to second order in the fields, where $v^i_{\pi , L}$ is the long wavelength mode generated by the spatial derivative of $\pi_L$, i.e.,
\be
\label{vipi}
v^i_{\pi , L} \equiv -a^{-1} \partial_i \pi_L  \;,
\ee
we find,
\begin{align} 
&a^{-2} \partial^2 \varphi_{a }^{(2),{\rm NL}}  \approx \nu_{\varphi_a} (a^{-1} v_{\pi,L}^i \partial_i \delta^{(1)} ) \\
&\quad   \quad + \sigma_{\varphi_a} a^{-1} \left( 2 v_{\pi,L}^i \partial_i \dot \delta^{(1)} + \dot v_{\pi,L}^i \partial_i \delta^{(1)} - H v_{\pi,L}^i \partial_i \delta^{(1)} \right)  \nonumber \\
& \quad  \quad =  -  \left(   \nu_{\varphi_a} L_\pi + \sigma_{\varphi_a} ( 3 H  f L_\pi + \dot L_\pi    )      \right)\frac{\partial_i \delta_{L}^{(1)}}{\partial^2} \partial_i \delta^{(1)}  \nonumber \ .
\end{align}
Here for the last equality we have used  eq.~\eqref{Ldefs1} to replace $v^i_{\pi , L}$ by its expression in terms of $\delta_L$, valid in the linear regime.  Comparing this expression with \eqn{phi2nllimit}, we see that the symmetry \eqn{newtransf1} forces
\be \label{chinalpha}
\upsilon^{\varphi_a}_\alpha =  -    \nu_{\varphi_a} L_\pi - \sigma_{\varphi_a} ( 3 H  f L_\pi + \dot L_\pi    )      \ .
\ee
This expression is in agreement with the full calculation presented in \appref{sec:coeff}.

The symmetry \eqn{transf1} allows us to easily determine otherwise complicated coefficients in the non-linear equations in terms of the coefficients in the linear equations. For instance, we see immediately why Horndeski theories do not generate terms proportional to $X_{\alpha_s} $: {there} are no time-derivatives in the field equations and thus no terms containing $\dot \delta$ and $\ddot \delta$  in \eqn{secondorderpoisson}. In \secref{symmetrysec} we will return to this symmetry and discuss its consequences on the full second-order solution for $\delta$  and the consistency relations.

%%%%%%%%%%%%%%%%%%%
%
%
%
%
\section{Fluid equations} \label{fluideqssec}

The equations governing the matter sector in the non-relativistic limit are the fluid equations 
\begin{align}
\begin{split} \label{eom1}
& \dot \delta + a^{-1} \partial_i \left( (1 + \delta) v^i \right) = 0 \ ,  \\
& \dot v^i + H v^i + a^{-1} v^j \partial_j v^i + a^{-1} \partial_i \Phi = 0   \ , 
\end{split}
\end{align}
where $v^i$ is the matter velocity.  In writing these equations we have assumed that matter is minimally coupled to the gravitational metric. Therefore, we work in the so-called Jordan frame.

Combining these two equations, we have, 
\be \label{secondordereom}
\ddot \delta + 2 H \dot \delta - a^{-2} \partial^2 \Phi = -a^{-2} \partial_i  \left( \partial_t ( a \delta v^i) -  v^j \partial_j v^i  \right) \ , 
\ee
and so we see that we need $\partial^2 \Phi$ in terms of $\delta$ from \secref{scalarfieldeqssec} to complete the system of equations.

%%%%%%%%%
%
\subsection{Perturbative solutions}

%%%%%%%%%%%%%%
\subsubsection{Linear solutions } \label{fluidseclin}

Using \eqn{linearsols} for $\partial^2 \Phi^{(1)}$, the linear equation for $\delta^{(1)}$ is 
\be \label{lineareom}
\ddot \delta^{(1)} + \bar \nu_\Phi \dot \delta^{(1)} - \bar \mu_\Phi \delta^{(1)} = 0 \ , 
\ee
where for future convenience, we have defined 
\be
\bar \nu_\Phi \equiv \frac{2 H - \nu_\Phi}{1 - \sigma_\Phi} \ , \quad \bar \mu_\Phi \equiv \frac{\mu_\Phi }{1 - \sigma_\Phi}  \ . 
\ee
The linear equation \eqn{lineareom} has two solutions, one growing, $D_+(t)$, and one decaying, $D_- (  t )$.  We focus on the growing mode solution, which will be used in the quadratic terms of the second-order equation, so we write the solution for $\delta^{(1)}$ as \eqn{lineardelta}.  Looking at \eqn{eom1}, this means that the linear solution for the velocity can be written
\be \label{linearvdef}
v^{(1)i} = - a \frac{ \partial_i \dot \delta^{(1)}}{\partial^2} = - a H f \frac{ \partial_i  \delta^{(1)}}{\partial^2}  \ . 
\ee
where $f$ is the linear growth rate defined in \eqn{fdef}.

%%%%%%%%
%
%
\subsubsection{Second-order solution}
Since we are interested in the second-order solution $\delta^{(2)}$ in this work, we can use the linear solutions $\delta^{(1)}$ and $v^{(1)i}$ in the quadratic terms in \eqn{secondordereom}.  Then, combining this with the expression for $\partial^2 \Phi^{(2)}$ from \eqn{secondorderpoisson}, we have the equation for the second-order field
\begin{align}
 & \hspace{-.05in} \ddot \delta^{(2)} + \bar \nu_\Phi \dot \delta^{(2)} - \bar \mu_\Phi \delta^{(2)} = \upsilon^\delta_\alpha X_{\alpha_s}  + \upsilon^\delta_\gamma X_\gamma     \ , 
\end{align}
where 
\begin{align} \label{upsilondeltaalpha}
\upsilon^\delta_\alpha &  = \frac{1}{1-\sigma_\Phi}  \left( 3 f^2 H^2 + H \dot f + f ( 2 H^2 + \dot H) + \upsilon^\Phi_\alpha \right)  \ ,  \nonumber \\
\upsilon^\delta_\gamma & = - \frac{1}{1-\sigma_\Phi}\left( f^2 H^2 - \upsilon^\Phi_\gamma \right)  \ . 
\end{align}
This means that the solution is
\begin{align} \label{delta2sol}
 \delta^{(2)} ( t ) = \int_0^t d t_1 \bar G( t , t_1 ) \left(  \upsilon^\delta_\alpha X_{\alpha_s}  + \upsilon^\delta_\gamma  X_\gamma    \right)_{t_1} 
\end{align}
where $\bar G( t , t_1 )$ is the Green's function, defined in \eqn{gfdef}.
Here, and in the rest of the paper, the subscript $t_1$ means that all time arguments inside of the brackets which are not explicitly shown are evaluated at $t_1$.

In Fourier space, \eqn{delta2sol} is
\begin{align}  
\delta^{(2)} ({\kvec}, t ) = \int^{\kvec}_{\kvec_1 , \kvec_2} F_2 ( \kvec_1 , \kvec_2 ; t )   \,  \delta^{(1)} ({\kvec_1},t ) \delta^{(1)} ({\kvec_2}, t) \ ,  \label{delta2sol2}
\end{align} 
where
\be \label{F2def}
 F_2 ( \kvec_1 , \kvec_2 ; t ) =  A_\alpha( t ) \alpha_s ( \kvec_1 , \kvec_2 ) + A_\gamma (  t ) \gamma ( \kvec_1 , \kvec_2 )  \ , 
\ee
and\footnote{ 
With the second-order solution for $\delta$ in \eqn{delta2sol2}, we can straightforwardly  compute the second-order solution for the velocity divergence $\theta \equiv a^{-1} \partial_i v^i $,
using the continuity equation \eqn{eom1} and the linear solution for $v^i$ \eqn{linearvdef}. 
In Fourier space, this becomes
\begin{align}  
\theta^{(2)} ({\kvec}, t ) = \int^{\kvec}_{\kvec_1 , \kvec_2} G_2 ( \kvec_1 , \kvec_2 ; t )   \,  \delta^{(1)} ({\kvec_1}, t ) \delta^{(1)} ( {\kvec_2}, t) \ ,  \label{theta2sol2}
\end{align} 
where (again suppressing time arguments)
\begin{align}
\begin{split}
G_2 ( \kvec_1 , \kvec_2  ) & = -(  \dot A_\alpha + 2 H f A_\alpha - H f ) \alpha_s ( \kvec_1 , \kvec_2 ) \\
& \quad \quad  -(\dot A_\gamma + 2 H f A_\gamma) \gamma ( \kvec_1 , \kvec_2 ) \ .
\end{split}
\end{align}
{For the implications of mass and momentum conservation for the velocity field, see \cite{Mercolli:2013bsa}.}}
\begin{align}
\begin{split} \label{a1and2}
A_\alpha(t ) & = \int_0^t d t_1 \bar G ( t , t_1 ) \upsilon_\alpha^\delta ( t_1 ) \frac{D_+(t_1)^2}{D_+ ( t)^2} \ , \\ 
A_\gamma (t) & = \int_0^t d t_1 \bar G ( t , t_1 ) \upsilon_\gamma^\delta ( t_1 ) \frac{D_+(t_1)^2}{D_+ ( t)^2} \ .
\end{split}
\end{align}

It is possible to further simplify the coefficient $A_\alpha ( t )$, as we show in \appref{app:green2}.  The result is 
\begin{align} \label{aalphaid}
A_\alpha ( t ) = 1  +  \int_0^t d t_1 \bar G( t , t_1 ) K_2 ( t_1 ) \frac{D_+(t_1)^2}{D_+(t)^2} \ , 
\end{align}
where
\be \label{k2def}
K_2 =   \frac{    \nu_\Phi L_{\Delta v}    +  \sigma_\Phi ( 3 H  f L_{\Delta v} + \dot L_{\Delta v}    )  }{1 - \sigma_\Phi  }     \ ,
\ee  
and
\be
L_{\Delta v} \equiv H f - L_\pi \ 
\ee
is defined analogously to \eqn{linearvdef} for the relative velocity 
\be
\label{Dv}
\Delta v^i \equiv v^i - v^i_{\pi} \ .
\ee
(We will return to this coefficient in the next subsection, in relation to symmetries of the field and fluid equations.) For Horndeski theories (which include the EdS (Einstein de Sitter) approximation and $\Lambda$CDM) we have $\nu_\Phi=\sigma_\Phi=0$ and thus $A_\alpha ( t ) = 1$. {In \secref{CRHorndeski} we will discuss how this value is fixed by the consistency relations in Horndeski theories.}  Only DHOST theories can change this coefficient. This was shown in \cite{Hirano:2018uar} restricting to GLPV theories.

 The coefficient $A_\gamma ( t ) $ has a complicated expression, in general.  It simplifies in the EdS approximation, where $A_\gamma (t) = -2/7$, but in $\Lambda$CDM and beyond it is in general time dependent \cite{Bernardeau:2001qr}. A study of this coefficient in Horndeski theories can be found in \cite{Bartolo:2013ws, Takushima:2013foa, Bellini:2015oua, Bellini:2015wfa,Burrage:2019afs}.

\begin{figure*}[t] 
\centering 
\hspace{-.3in} \includegraphics[width=0.475\textwidth]{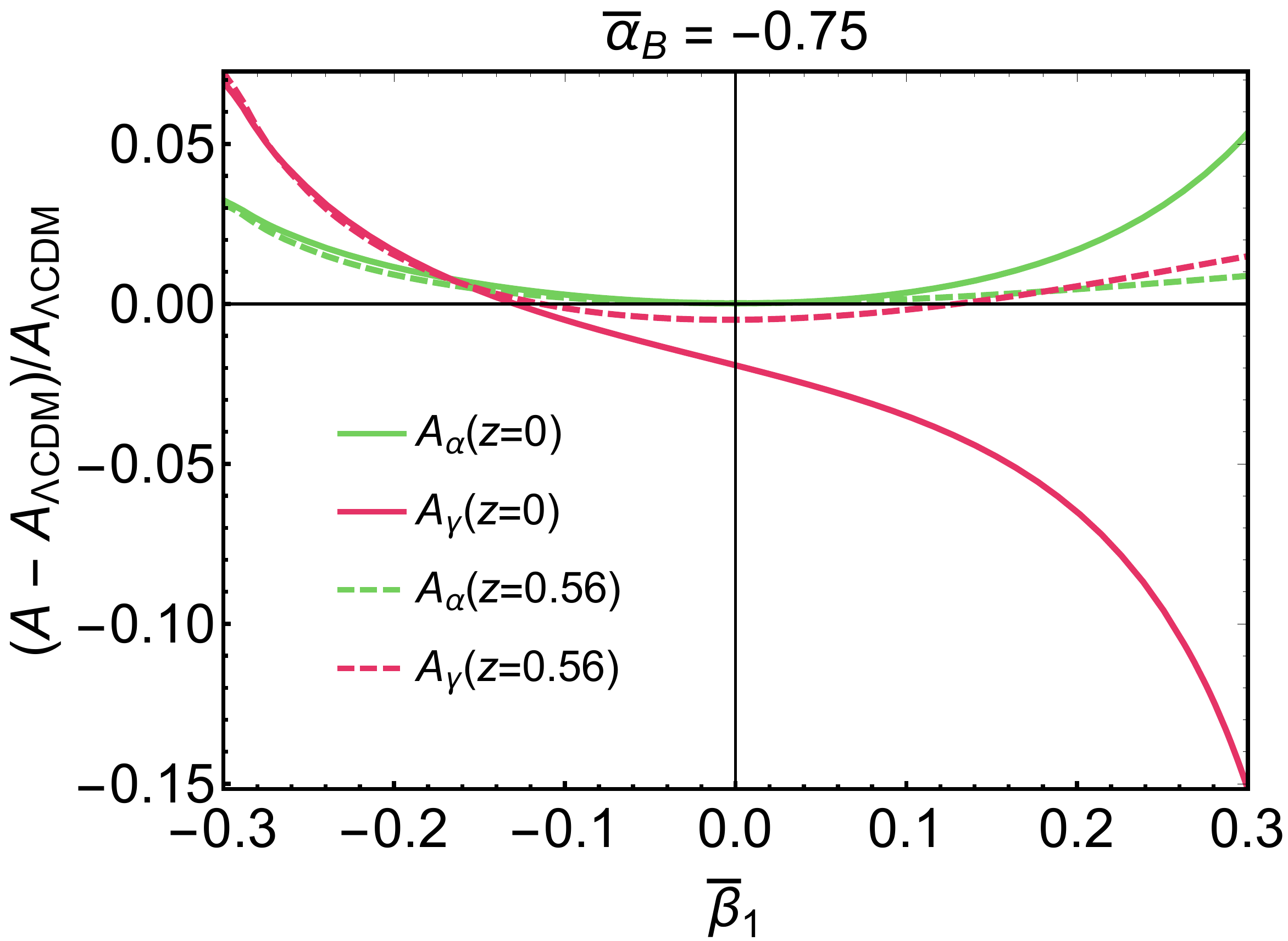} \hspace{0.2em}
\includegraphics[width=0.47\textwidth]{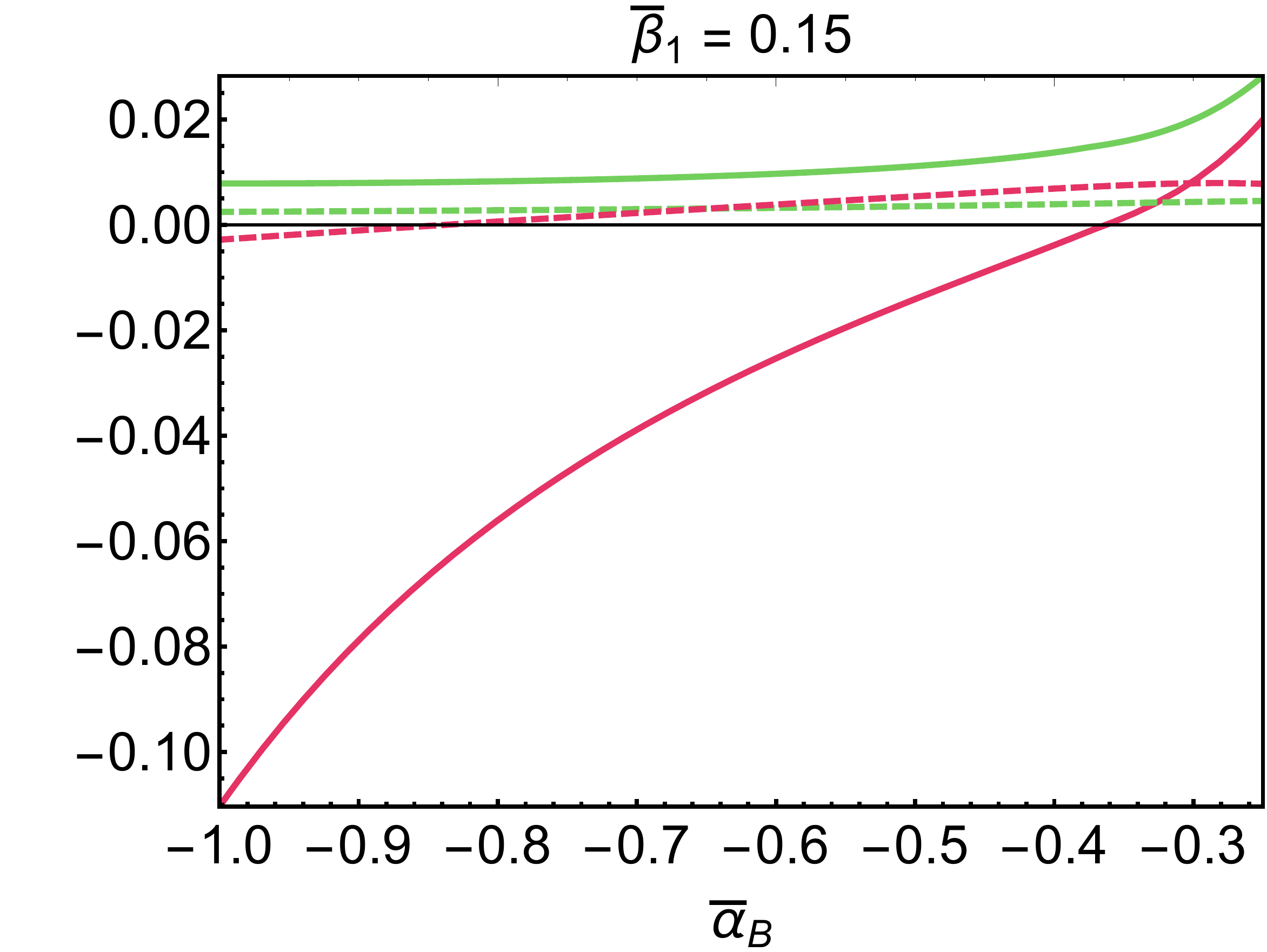}
\caption{ Functions $A_\alpha$ and $A_\gamma$, see \eqn{a1and2},  for two values of redshift, $z=0$ and {$z=0.56$}  as a function of the EFT of dark energy parameters $\alphaB$ and $\beta_1$.  
Specifically, we plot their fractional difference  from their $\Lambda$CDM values ($A_\alpha = 1$ and $A_\gamma = -0.284$ for $z=0$ and $A_\alpha = 1$ and $A_\gamma = -0.285$ for {$z=0.56$}).  
The background evolution has been chosen to be the one of $\Lambda$CDM, i.e.~the Hubble rate is given by $H ( a )  = H_0 \sqrt{a^{-3} \Omega_{\rm m , 0} + 1 - \Omega_{\rm m , 0}} $, the matter evolution is given by {$\Omega_{\rm m} ( a )   = \Omega_{\rm m , 0} / ( \Omega_{\rm m , 0} + a^3 ( 1 - \Omega_{\rm m , 0} )) $}, and we have taken $\Omega_{\rm m,0} = 0.281$ as the current value of the fractional matter density. (In the numerical calculation, the Hubble rate always appears in the combination $H/H_0$ so that the plots are independent of the value of $H_0$.)  
We  parametrize the time dependence of the  EFT parameters as $\alphaB ( a ) = \bar \alpha_{\rm B} ( 1 - \Omega_{\rm m} ( a ) )$ and $\beta_1 ( a ) = \bar \beta_{1} ( 1 - \Omega_{\rm m} ( a ) )$, where $ \bar \alpha_{\rm B}$ and $\bar \beta_1$ are constants (see e.g.~\cite{Bellini:2014fua}).
The other EFT parameters are chosen such that the model leaves the gravitational wave speed, amplitude, and decay unaffected (see e.g.~\cite{Crisostomi:2019yfo} for a discussion), i.e.~$\alphaT=\alphaM = 0$ and $\alphaH = - 2 \beta_1$.
Moreover, we  only plot values of $\bar \alpha_{\rm B}$ and $\bar \beta_{1}$ for which $ \alpha c_s^2 > 0$, as required by the absence of ghost and gradient instability (see e.g.~\cite{Gleyzes:2014rba}).  On the right-hand  {panel,  the case $\bar \alpha_{\rm B} <-1$ has $ \alpha c_s^2 > 0$, but since $A_\gamma$ becomes very large in this case,}  we do not plot the corresponding range.
{Notice that $A_\alpha = 1 $ in Horndeski theories, i.e.~for $\beta_1 = 0$, as expected. }    } \label{Asfigure}
\end{figure*}
We plot these functions for two different redshifts and different values of the EFT parameters in \figref{Asfigure}. As expected, $K_2=0$ and thus $A_\alpha=1$ for Horndeski theories ($\beta_1=0$) while $A_\gamma$ is modified in both Horndeski and DHOST theories.

Notice that we can organize the kernel in \eqn{F2def} as a multipolar expansion in the angle $\mu \equiv \hat k_1 \cdot \hat k_2$, i.e. in terms of the monopole (proportional to $\mu^0$), dipole (proportional to $\mu^1$)
and quadrupole (proportional to $\mu^2 - 1/3$) contributions.  Explicitly, we have (suppressing the time argument)
\begin{align}
\begin{split} \label{multipoles}
 F_2 ( \kvec_1 , \kvec_2  ) & = A_\alpha  + \frac23 A_\gamma  + A_\alpha \frac{\mu}{2} \left( \frac{k_2}{k_1} + \frac{k_1}{k_2} \right) \\
&  \quad    - A_\gamma \left( \mu^2 - 1/3  \right) \ .
 \end{split}
\end{align}

As expected, the solution \eqn{delta2sol} respects the conservation of mass and momentum,  since\footnote{As discussed in \cite{peebles}, this means that in Fourier space, 
\be
\delta^{(2)}( \kvec) \propto k^2
\ee
for $k \rightarrow 0$, which one can explicitly verify for the solution \eqn{delta2sol}. This contributes to the power spectrum with a term $\propto k^4$, which is why one includes the so-called stochastic contribution in the EFT of LSS \cite{Carrasco:2012cv}.} 
\be
\int d^3 x \, \delta^{(2)}(  \xvec  , t )  = 0  \quad \text{and} \quad \int d^3 x \, x^i \delta^{(2)} ( \xvec , t ) = 0 \ . 
\ee
In fact, mass and momentum conservation is the reason that the non-linear corrections in \eqn{delta2sol} appear in the specific combinations \eqn{xdefs}.

%%%%%%%%%%%
%
%
\subsection{Symmetries of the fluid equations and infrared behavior}
\label{symmIR}

To find the leading terms in the IR limit, one could of course start with the explicit solution \eqn{delta2sol} and take the IR limit.  However, we are going to show that the leading IR behavior is related to the symmetries of the gravitational field equations, discussed in \secref{leadingirtermssec}, and the symmetries of the fluid equations, which we discuss next.

The fluid equations \eqn{eom1} are invariant under the following coordinate change and shifts of the fields:
\begin{align}
\begin{split}  \label{transf2}
 \tilde x^i &= x^i +  n^i ( t )  \ , \quad {\tilde t  = t }\ ,  \\
 \tilde \varphi_a ( \tilde x^j , t )  & = \varphi_a ( x^j , t)  +h^i_{\varphi_a} ( t ) \tilde x^i  \ , \\
      \tilde \delta ( \tilde x^j , t ) &= \delta ( x^j , t ) \ , \\
 \tilde v^i ( \tilde x^j , t ) & = v^i( x^j , t) + a \dot n^i ( t )  \ ,
\end{split}
\end{align}
for generic $n^i(t)$ and $h^i_{\Psi, \pi} ( t )$, as long as 
\be
h_{\Phi}^i(t) = - a^2( \ddot n^i(t) + 2 H \dot n^i(t) ) \ .
\ee
These symmetries have been discussed to derive the consistency relations of LSS, in e.g.~\cite{Scoccimarro:1995if,Peloso:2013zw,Kehagias:2013yd,Creminelli:2013mca,Horn:2014rta}, where they apply to both the fluid and gravitational field equations. { Here, we have introduced different notation from the transformation \eqn{transf1} to facilitate our discussion of the adiabatic mode construction in \secref{symmetrysec}. }

Equivalently, the fluid equations are invariant under the replacements 
\begin{align} \label{fluidreps}
\partial_i & \rightarrow \partial_i \ , \quad  \partial_t   \rightarrow \partial_t - \dot n^i ( t )   \partial_i  \ , \\ 
\varphi_a  & \rightarrow \varphi_a + h^i_{\varphi_a}(t) x^i \ , \quad  \delta  \rightarrow \delta \ ,  \quad  v^i  \rightarrow v^i +  a \dot n^i ( t )    \  . \nonumber
\end{align}
One can explicitly check that the leading IR terms on the right-hand side of \eqn{secondordereom} are generated by this transformation.  

In the gravitational equations, the transformation of $\pi$, \eqn{bipi}, is related to the coordinate change (while the transformations of the other fields are arbitrary). In the fluid equations, the transformations of $v^i$ and $\Phi$ are related to the coordinate transformation (while the transformations of the other fields are arbitrary).  {  These transformations can be combined by taking $\xi^i(t) = n^i(t)$, $b^i_{\Phi,\Psi} (t) = h^i_{\Phi,\Psi} (t)$, and  $b^i_\pi ( t ) = -a^2 \dot n^i(t)$
to give an overall Galilean invariance.  In this case, by \eqn{transf1}, \eqn{vipi}, and \eqn{fluidreps}, the transformations of $v^i$ and $v^i_\pi$ are the same, so both velocities can only be simultaneously eliminated if there is no relative velocity on large scales.  As we will see later, this means that a physical adiabatic mode cannot be constructed.}

Now, we show how these symmetries determine the leading IR behavior of $\delta^{(2)}$.  We start with the equations for the fluid and the gravitational sector separately.  Taking the IR limit of \eqn{secondordereom} (or equivalently using the transformations \eqn{fluidreps}), we have
\begin{align} \label{ireq1}
& \ddot \delta^{(2)}  + 2 H \dot \delta^{(2)}  - a^{-2} \partial^2 \Phi^{(2)}  \\
& \qquad   \approx -a^{-1}  \left( 2 v_{L}^i \partial_i \dot \delta^{(1)} + \dot v_{L}^i \partial_i \delta^{(1)} +  H v_{L}^i \partial_i \delta^{(1)} \right) \ . \nonumber
\end{align}
The leading IR non-linear terms which must be present in the non-linear extension of \eqn{linearsols} can be obtained using \eqn{deltadottrans}, giving 
\be
\begin{split} \label{ireq2}
& a^{-2} \partial^2 \Phi^{(2)}    \approx  \mu_\Phi \delta^{(2)}  + \nu_\Phi \dot \delta^{(2)}  + \sigma_\Phi \ddot \delta^{(2)}   + \nu_{\Phi} a^{-1} v_{\pi,L}^i \partial_i \delta^{(1)}   \\
& \quad   + \sigma_{\Phi} a^{-1} \left( 2 v_{\pi,L}^i \partial_i \dot \delta^{(1)} + \dot v_{\pi,L}^i \partial_i \delta^{(1)} - H v_{\pi,L}^i \partial_i \delta^{(1)} \right) \ .   
\end{split}
\ee

Now, we combine \eqn{ireq1} and \eqn{ireq2}, divide by $1 - \sigma_\Phi$, and simplify the expression further to write 
\begin{align} \label{delta2ireq}
& \ddot \delta^{(2)}  + \bar \nu_\Phi \dot \delta^{(2)} - \bar \mu_\Phi \delta^{(2)}  \approx   2 \frac{\partial_i \dot \delta }{\partial^2} \partial_i \dot \delta  + \frac{\partial_i \ddot \delta}{\partial^2 } \partial_i \delta  + \bar \nu_\Phi  \frac{\partial_i \dot \delta}{ \partial^2 } \partial_i \delta \nonumber \\
& \hspace{.1in} - \frac{\sigma_\Phi a^{-1}}{1-\sigma_\Phi} \left( 2 \Delta v^i \partial_i \dot \delta + \Delta \dot v^i \partial_i \delta - H \Delta v^i \partial_i \delta \right) \nonumber \\
& \hspace{.1in} -  \frac{ \nu_\Phi a^{-1} }{1-\sigma_\Phi}   \Delta v^i \partial_i \delta  \ ,
\end{align}
where it is understood that all of the fields on the right-hand side are the linear fields.  

Now, to solve for $\delta^{(2)}$ in the squeezed limit, we apply the Green's function.  Explicit details are given in \appref{app:green2}.  We obtain,
\begin{align}
\begin{split} \label{delta2irsol}
\delta^{(2)} ( t ) & \approx  \frac{ \partial_i  \delta^{(1)} ( t)}{\partial^2}  { \partial_i } \delta^{(1)} ( t )  \\
& + \int_0^t d t_1 \bar G( t , t_1 ) K_2 ( t_1 )   \frac{\partial_i \delta^{(1)} ( \ttt ) }{\partial^2 } \partial_i \delta^{(1)} ( \ttt ) \ ,
\end{split}
\end{align}
where $K_2 ( t )$ was defined in \eqn{k2def}.  Upon Fourier transforming the above, and comparing it with the squeezed limit of the full solution for $\delta^{(2)}$ in \eqn{delta2sol2}, we can verify that this result agrees with that of \eqn{aalphaid} in the squeezed limit.

Let us make a few comments about the solution \eqn{delta2irsol}.  Firstly, this shows again that the dipole term is in general modified in GLPV theories \cite{Hirano:2018uar}, and that this happens also for the more general DHOST theories.  Secondly, the construction in this section allows us to see explicitly how the change in the dipole is determined by the coefficients of the linear solutions, i.e. $\nu_\Phi$ and $\sigma_\Phi$.  Finally, we also see explicitly that the change in the dipole is proportional to the relative velocity $\Delta v^i$, or equivalently $L_{\Delta v}$.  We will comment more on this last point in the next section.

%%%%%%%%%%%%%%%
%
%
%
%
\section{Symmetries and consistency relations} \label{symmetrysec}

In the previous sections we have discussed two different coordinate and field transformations: \eqn{transf1} and \eqn{transf2}. The scalar field equations, \eqns{phieq1}{pieq1}, are invariant under the former and the fluid equations, \eqn{eom1}, are invariant  under the latter.   

 We will now discuss the consequences of these { transformations} on the correlation functions of the density contrast, distinguishing between two cases: Horndeski theories, { where the large-scale velocity can be removed by a coordinate transformation, and DHOST theories, where there are two large-scale velocities which, because of eq.~\eqref{bipi}, cannot both be simultaneously eliminated.}

%%%%%%%%%%%%
\subsection{Horndeski theories}
\label{CRHorndeski}

For Horndeski theories (which include $\Lambda$CDM) we can set 
\be
\label{equal}
\xi^i(t) = n^i(t) \;, \qquad b^i_{\varphi_a} (t) = h^i_{\varphi_a} (t)  
\ee
in eq.~\eqref{transf1}, so that the full field and fluid equations are invariant under the same transformations, \eqn{transf2}. We can then use the invariance of the equations under these transformations to derive  the so-called consistency relations for LSS valid in $\Lambda$CDM \cite{Peloso:2013zw,Kehagias:2013yd,Creminelli:2013mca,Horn:2014rta}.\footnote{The proof of the consistency relations relies on Gaussian initial conditions, so that there is no correlation between long and short modes in the initial state.  We shall assume this throughout this paper.}

A way to derive these relations  is by using the adiabatic mode construction  \cite{Weinberg:2003sw}, that we extend here to Horndeski theories.  
Starting from some solution to   the gravitational and fluid equations, \eqns{phieq1}{pieq1} and \eqref{eom1}, 
the transformation \eqn{transf2}  generates a new solution for any $n^i(t)$, $h_\Psi^i (t )$, and $h_\pi^i ( t )$. In this way, we can derive the effect of a long wavelength mode on a short one, at least locally, and  determine the statistical properties of the density field in the squeezed limit. 
%Note that this construction is based on the equivalence principle: the effect of a long wavelength mode can be locally erased by a change of coordinates \cite{Creminelli:2013mca}.

However, in order to ensure that the long mode is the small momentum limit of a real physical solution, we need to verify that it satisfies the equations of motion that vanish in the small momentum limit, i.e. that have enough spatial derivatives to make the transformation \eqn{transf2} trivially {a symmetry}. 

First of all, to have a physical solution with a particular long wavelength velocity $v^i_L$, we choose $\dot n^i_L = a^{-1} v^i_L$ (we have included the subscript $L$ on $n^i_L$ to stress that we are giving $n^i$ a very weak spatial dependence).  Then, using the linear continuity equation, we see that 
\be
\label{deltaL}
\delta_L = - \partial_i n^i_L\;, 
\ee
and plugging this into the Euler equation, we have 
\be
-\partial_i \left( \ddot n^i_L + 2 H \dot n^i_L - \mu_\Phi n^i_L \right) = 0 \ .
\ee
Of course, if $n^i_L$ only depends on time, this equation is trivially satisfied.  However, in order to have a physical solution, we demand that the equation is satisfied after removing the spatial derivative, i.e. that
\be \label{nieq}
\ddot n^i_L + 2 H \dot n^i_L - \mu_\Phi n^i_L = 0 \ . 
\ee

Now, we move to the modified Poisson equations, {which in Horndeski theories take the form} $a^{-2} \partial^2 \varphi_a = \mu_{\varphi_a} \delta$.  Again, these equations are trivially satisfied for purely time dependent $n^i$, and $h^i_{\varphi_a}$, but we demand that they are satisfied with one derivative stripped off, i.e.
\be \label{bieq}
a^{-2} h^i_{\varphi_a , L} = - \mu_{\varphi_a} n^i_L \ . 
\ee
Because it is possible to choose the time dependence such that \eqn{nieq} and \eqn{bieq} are satisfied (these equations are equivalent to the linear equations of motion), this means that we can successfully generate a physical adiabatic mode which can be used to derive the consistency relations.   

The same logic also explains why the dipole term is not changed in Horndeski theories. At least locally, one can use the transformations \eqref{transf2} to boost to the rest frame of the matter and remove the long-wavelength matter velocity $v_L^i$, by $x^i \rightarrow x^i + n_L^i$, where $v^i_L = a \dot n_L^i$. {In Fourier space, the effect of this boost on a short mode is 
\be
\begin{split}
\delta({\vec k}) \rightarrow \delta({\vec k}) e^{-i \vec k \cdot \vec n_L} &  \simeq  \delta({\vec k}) + \int_{\kvec_1 , \kvec_2}^{\kvec}  \frac{\kvec_1 \cdot \kvec_2 }{k_1^2}  \delta(\kvec_1  ) \delta(\kvec_2) \;,
\end{split}
\ee
where we have expanded the exponential to first order and used the continuity equation in the form \eqn{deltaL}.  {This matches the Fourier transform of the squeezed limit expression \eqn{delta2irsol} in the case of Horndeski, i.e.~for $K_2(t) = 0$.}

\subsection{DHOST theories} \label{dhostsymm}
If the relative velocity $\Delta v^i \neq 0$, the adiabatic mode construction discussed above does not apply. { This is because one cannot remove both large-scale velocities with a single transformation of the form \eqn{transf2}.}  { Therefore}, we expect  the consistency relations  to be generically violated.

To  see that $\Delta v^i$ cannot be removed by a coordinate transformation, imagine imposing \eqn{equal} and doing the transformation \eqn{transf2} to try to generate a physical adiabatic mode.  In order to try to enforce that the scalar field equations are also invariant under \eqn{transf2}, we would need to take $h^i_{\pi,L} = -  a^2 \dot n^i_L$.  Then, the linear equation for $\partial^2 \pi$ \eqn{linearsols} would demand 
\be \label{bheq1}
\dot n^i_L = \mu_\pi n^i_L + \nu_\pi \dot n^i_L
\ee
whereas the linear equation for $\delta$ demands
\be \label{bheq2}
(1 - \sigma_\Phi ) \ddot n^i_L + ( 2 H - \nu_\Phi ) \dot n^i_L - \mu_\Phi n^i_L = 0 \ . 
\ee
For generic time-dependent coefficients, it is not possible to simultaneously solve \eqn{bheq1} and \eqn{bheq2}, unless, for example, \eqn{bheq1} becomes trivial by having $H f = \mu_\pi + H f \nu_\pi$, which is simply the condition that $L_{\Delta v} = 0$, i.e. that the relative velocity in eq.~\eqref{Dv} vanishes.

The origin of this effect, absent in Horndeski theories, lies on the kinetic coupling between matter and the scalar field, also called kinetic matter mixing. As discussed in \cite{DAmico:2016ntq}, in theories beyond Horndeski matter is kinetically mixed with the scalar field. The effect of a time dependent boost generates a long-wavelength mode of $\pi$ affecting this mixing. Since the velocity of $\pi$, $v_\pi^i$ and that of the fluid $v^i$ are generally different,   one cannot simultaneously remove the kinetic matter mixing and the convective motion of the fluid by a single boost.

Because the consistency relation can be violated, we find that the dipole term \eqn{delta2irsol} can also be changed from the standard, single-velocity case, which generalizes the analysis of \cite{Hirano:2018uar} which restricted to GLPV theories.  Although the  consistency relations are violated, the symmetries \eqref{transf1} and \eqref{transf2}  allow us to universally determine the dipole \eqn{delta2irsol} in terms of the coefficients of the linear equations $\nu_\Phi$ and $\sigma_\Phi$, as shown in Sec.~\ref{symmIR}.  The deviation is proportional to the relative velocity $L_{\Delta v}$, as expected.  

 { Although we only have a single dynamical fluid here,} the current situation is similar to the case of multiple fluids, like dark matter and baryons, which have a non-zero relative velocity \cite{Bernardeau:2012aq,Peloso:2013spa}.

%%%%%%%%%%%%%%%%%%%%%%%%%%%%%%%%%%%%%%%%%%%%%%%%%%%%%%%%%%%%%%%%
%%%%%%%%%%%%%%%%%%%%%%%%%%%%%%%%%%%%%%%%%%%%%%%%%%%%%%%%%%%%%%%%
%%%%%%%%%%%%%%%%%%%%%%%%%%%%%%%%%%%%%%%%%%%%%%%%%%%%%%%%%%%%%%%%
%%%%%%%%%%%%%%%%%%%%%%%%%%%%%%%%%%%%%%%%%%%%%%%%%%%%%%%%%%%%%%%%

\section{Bispectra} \label{bispectrumsec}

{Here we explore the observational consequences of what {was} discussed in the previous sections on the tree-level bispectra of the cosmic fields. Our main results rely on \eqn{aalphaid}, that $A_\alpha \neq 1$ in general in DHOST theories.  This changes the $k_1 / k \rightarrow 0$ limit of the second order field \eqn{delta2sol2}, which in turn changes the squeezed limit of the bispectrum, as we show next, from its universal value in $\Lambda$CDM, Horndeski, and other theories that satisfy the consistency relations.  We start correlating the same field, i.e.~the density contrast.}

\subsection{Auto-correlation}
\label{bispectrumsec1}
We can use the perturbative  calculations of Sec.~\ref{fluideqssec} to compute the equal-time bispectrum of $\delta$, $B(k_1 , k_2 , k_3)$, defined by 
\be \label{bispectrumdef}
\langle \delta({\kvec_1}) \delta({\kvec_2}) \delta({\kvec_3}) \rangle = ( 2 \pi )^3 \delta_D ( \kvec_1 + \kvec_2 + \kvec_3 )B ( k_1 , k_2 , k_3) \ . 
\ee
Expanding $\delta = \delta^{(1)} + \delta^{(2)} + \ldots$, using the explicit solution for $\delta^{(2)}$ in \eqn{delta2sol2} and assuming Gaussian initial conditions, 
we have, at tree level,
\be
B ( k_1 , k_2 , k_3 ) = 2  F_2 ( \kvec_1 , \kvec_2 ) P_{11} ( k_1) P_{11} ( k_2 ) + ( 2 \text{ perms.}) \;,
\ee
where $P_{11}(k)$ is the linear power spectrum of $\delta$, defined by
\be \label{PSdef}
\langle \delta^{(1)} ({\kvec}) \delta^{(1)} ({\kvec'}) \rangle = ( 2 \pi )^3 \delta_D ( \kvec + \kvec' )P_{11} ( k) \ . 
\ee

\begin{figure*}[t] 
\centering 
\hspace{-.3in} \includegraphics[width=1\textwidth]{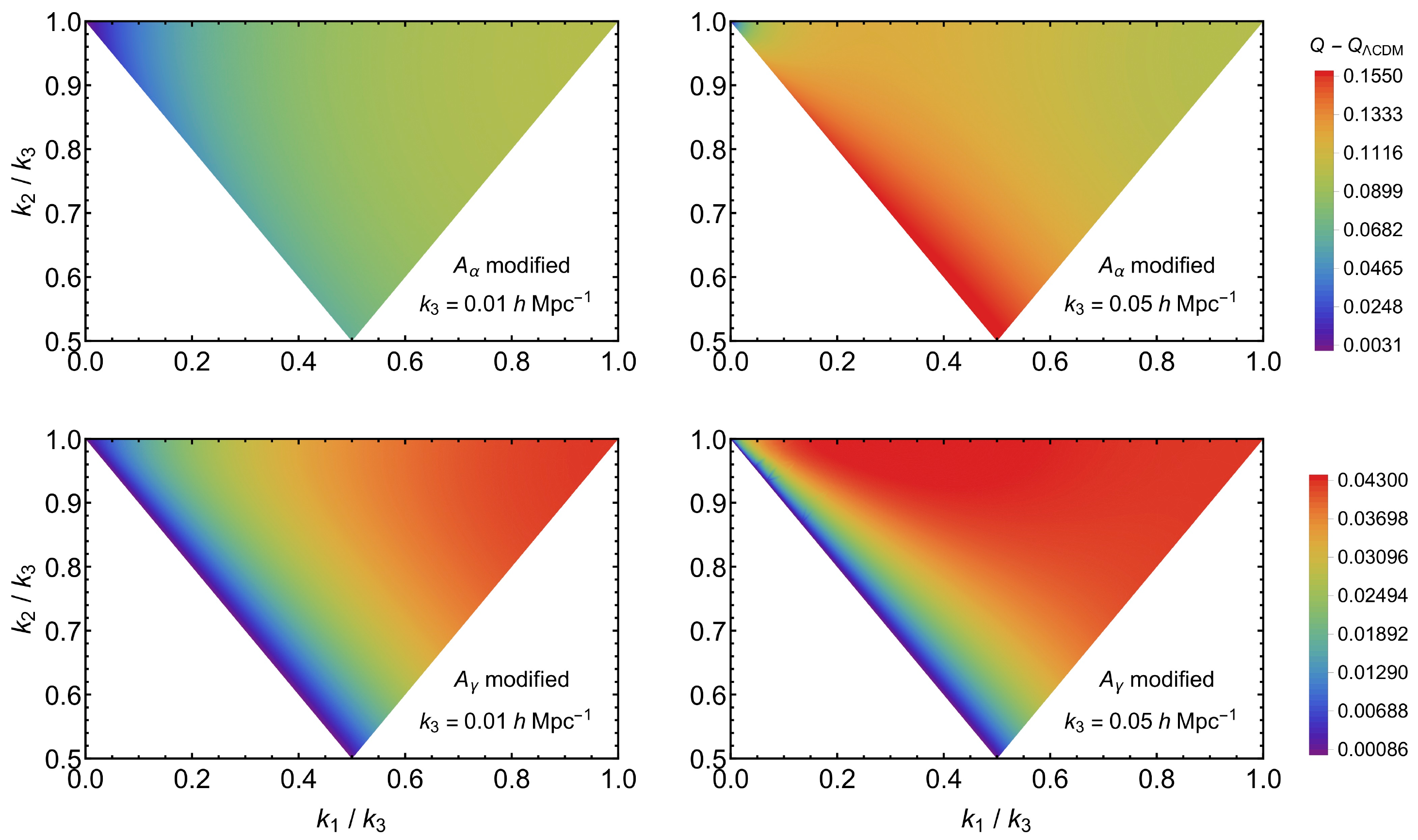}
\caption{ {Shape of the difference of the reduced bispectrum and the one in $\Lambda$CDM,  $Q ( k_1 , k_2 , k_3 )- Q_{\Lambda \text{CDM}} ( k_1 , k_2 , k_3 ) $, for two $10\%$ modifications away from {$\Lambda$CDM at $z=0$ (which has }$(A_\alpha, A_\gamma)= (1,-0.284)$, see Fig.~\ref{Asfigure}), i.e.~for $(A_\alpha,A_\gamma) = (1.1,-0.284)$ (upper panels) and $(A_\alpha,A_\gamma)  = (1,-0.256)$  (lower panels).   Only modifying $A_\alpha$ produces a signal for folded triangles (i.e. along the diagonal going from $(0,1)$ to $(0.5,0.5)$). As expected, the bispectrum {is not enhanced} in the squeezed limit (upper-left corner {of each plot}). }}
\label{fig:Fig1}
\end{figure*}

{In Fig.~\ref{fig:Fig1}, we plot the relative difference between  the amplitude of the reduced bispectrum, defined as
\be
Q ( k_1 , k_2 , k_3 ) \equiv \frac{B ( k_1 , k_2 , k_3 )}{P_{11} ( k_1) P_{11} ( k_2 ) + ( 2 \text{ perms.})} \;,
\ee
and the one of the reduced bispectrum in $\Lambda$CDM. Following \cite{Jeong:2009vd}, we plot this  as a function of the shape of the triangle formed by $(\kvec_1, \kvec_2, \kvec_3)$ with the condition $k_1 \le k_2 \le k_3$, for two values of $k_3$, i.e.~$k_3=0.01 \, h\text{Mpc}^{-1}$ and $k_3=0.05 \, h\text{Mpc}^{-1}$. To show the effect of DHOST theories, in the upper panels we plot this difference in the case where $A_\alpha$ is modified by $10\%$ from its $\Lambda$CDM value while $A_\gamma$ is unmodified. For comparison with more general modifications one can have in both Horndeski and DHOST theories, in the lower panels we consider the case where $A_\gamma$ is modified by $10\%$ from its $\Lambda$CDM value while $A_\alpha=1$. 

Changing either $A_\alpha$ or $A_\gamma$ modifies the {reduced} bispectrum for equilateral triangles (upper-right corner {of each plot}). However, as first noticed in \cite{Hirano:2018uar} a change in $A_\gamma$ does not produce any modifications for folded triangles $k_1 + k_2 = k_3$ (i.e. along the diagonal going from $(0,1)$ to $(0.5,0.5)$). Therefore, modifications of the bispectrum for folded triangles are unique signatures of DHOST theories.

There are no {enhanced} modifications in the squeezed limit (upper-left corner of each plot). Indeed,  {the leading contribution to the bispectrum} vanishes in this limit  for all cases.  This can be seen by 
%Let us consider the squeezed limit of the bispectrum. In particular, we 
defining $\vec q \equiv - \kvec_1$, $\kvec \equiv  \kvec_2 - \qvec/2$ and expanding} \eqn{bispectrumdef}, assuming $q \ll k$. The term of the bispectrum proportional to $F_2 ( \kvec_2 , \kvec_3 )$ can be neglected and the bispectrum gives, up to corrections of order ${\cal O}((q/k)^0)$,\footnote{We are assuming that the long mode is longer than the BAO scale of the baryon acoustic peak \cite{Baldauf:2015xfa}.} 
\begin{align}
\begin{split} \label{bispecsl}
\lim_{q \to 0} \frac{B (q, k_2, k_3)}{P_{11} ( q ) P_{11} ( k) }  \approx - 2  A_\alpha \left( \half \frac{\qvec \cdot \kvec}{q^2} - \half \frac{\qvec \cdot \kvec}{q^2} \right) {= } 0 \ .
\end{split}
\end{align}
Therefore, there is no $k/q$ enhancement in the squeezed limit $q \to 0$.  This would seem to suggest that the consistency relations are satisfied \cite{Peloso:2013zw,Kehagias:2013yd,Creminelli:2013mca,Horn:2014rta}.  However, the vanishing of the right-hand side of \eqn{bispecsl} is not a consequence of the consistency relations  but simply of the symmetry of the bispectrum under exchange of the two arguments $k_2$ and $k_3$ {(and translation invariance, i.e. that $\kvec_1 + \kvec_2 + \kvec_3 = 0$).}
Therefore, the violation of the consistency relation has no effect on the bispectrum computed from the auto-correlation. In order to see some effect in the 3-point function, we must  correlate different tracers \cite{Creminelli:2013nua}, as we do in the next subsection.

\subsection{Cross-correlation with the lensing potential}
\label{sec:lensing}

To see an enhanced effect in the squeezed limit, we need to consider correlations with different tracers, so that the bispectrum is no longer symmetric under the exchange of $\kvec_2$ and $\kvec_3$.  Following \cite{Creminelli:2013nua}, we can estimate the constraining power of a galaxy survey on this effect by considering the tree-level bispectrum involving the density contrast of two classes of objects  $A$ and $B$ (e.g.~galaxies with different masses). Defining
\be
\begin{split}
\langle \delta({\kvec_1}) \delta_{ A}({\kvec_2}) \delta_{ B}({\kvec_3}) \rangle= ( 2 \pi )^3 & \delta_D ( \kvec_1 + \kvec_2 + \kvec_3 )  \\
& \times B^{{ AB}} ( k_1 , k_2 , k_3 ) \;,
\end{split}
\ee
one expects a violation of the consistency relation of the type 
\be \label{squeezemml}
\lim_{q \to 0} \frac{B^{{ AB}} (q,k_2,k_3)}{P_{11} ( q ) P^{AB}_{11} ( k) } \approx  \epsilon \;  \frac{\qvec \cdot \kvec}{q^2} \; ,
 \ee
 where $P^{AB}_{11} ( k)$ is the linear cross-power spectrum between the two species and $\epsilon$ is a parameter that is non-vanishing only in theories beyond Horndeski. (This estimate holds also in redshift-space \cite{Creminelli:2013poa}.) For a (Euclid-like) survey with effective volume of $V_{\rm eff} \simeq 20  (\text{Gpc}/h)^{3}$, one expects a limit $\epsilon \lesssim 10^{-3}$ \cite{Creminelli:2013nua}.

As a simplifying, calculable example, we consider the 3-point correlation function $\langle \delta({\kvec_1}) \delta({\kvec_2}) \delta_{\rm lens}({\kvec_3}) \rangle$, where  $\delta_{\rm lens}$ is the ``lensing density,'' defined as
\be
\label{lenspot}
\delta_{\rm lens} \equiv ( 3 \Omega_{\rm m} H^2)^{-1} a^{-2} \partial^2\left( \Phi + \Psi \right) \ ,
\ee
where $\Omega_{\rm m} \equiv \bar \rho_{\rm m} / ( 3 H^2 M^2)$. Here $( \Phi + \Psi )/2$ is the so-called lensing potential,  which enters measurements of weak lensing convergence and shear (see for instance \cite{Lewis:2006fu}).
It is not directly an observable, but  lensing observables are built from projecting this quantity along the line of sight with some window function.

We want to compute the tree-level matter-matter-lensing bispectrum, defined by 
\be
\begin{split}
\langle \delta({\kvec_1}) \delta({\kvec_2}) \delta_{\rm lens}({\kvec_3}) \rangle= ( 2 \pi )^3 & \delta_D ( \kvec_1 + \kvec_2 + \kvec_3 )  \\
& \times B^{{\rm mm l}} ( k_1 , k_2 , k_3 ) \;.
\end{split}
\ee
As usual, we expand $\delta_{{\rm lens}}$ into first- and second-order parts.
From \eqn{Ldefs1} we have
\be
\delta^{(1)}_{{\rm lens}}( \kvec) = L_{\rm lens} \delta^{(1)}({\kvec} ) \ , 
\ee
where 
\be
\label{Llens}
L_{\rm lens} \equiv ( 3 \Omega_{\rm m} H^{2})^{-1} \left( L_\Phi + L_\Psi \right)\;.
\ee  
Next, 
we need $\delta_{\rm lens}$ at second order.
{Plugging $\delta^{(2)}$ and $a^{-2} \partial^2 \varphi_a^{(2),{\rm NL}} $, using \eqn{delta2sol2} and \eqn{secondorderform}, into the second-order Poisson equation} \eqn{secondorderpoisson}, and using this equation in \eqn{lenspot} above, we obtain
\begin{align} 
\delta^{(2)}_{{\rm lens}}(\kvec)  = \int^{\kvec}_{\kvec_1 , \kvec_2} F^{\rm lens}_2 ( \kvec_1 , \kvec_2 )   \,  \delta^{(1)}({\kvec_1})  \delta^{(1)}({\kvec_2})  \ ,  \label{delta2lenssol2}
\end{align} 
where
\begin{align}
\begin{split}
F^{\rm lens}_2 ( \kvec_1 , \kvec_2  ) & = A^{\rm lens}_{\alpha} \alpha_s ( \kvec_1 , \kvec_2 ) + A^{\rm lens}_{\gamma} \gamma ( \kvec_1 , \kvec_2 )  \ ,
\end{split}
\end{align}
with
\begin{align}
\label{Alens}
A^{\rm lens}_{\alpha} &=  (3 \Omega_{\rm m} H^2 )^{-1} \bigg[ \upsilon_{\alpha_s}^{\Phi} +\upsilon_{\alpha_s}^{\Psi} + (\mu_{\Phi} +\mu_{\Psi}) A_{\alpha}  \\
&+  \frac{(\nu_{\Phi} +\nu_{\Psi} ) \partial_t ( A_{\alpha} D_+^2) + ( \sigma_{\Phi}+\sigma_{\Psi}) \partial^2_t ( A_{\alpha} D_+^2)}{D_+^2}  \bigg]\;,   \nonumber
\end{align}
and an analogous expression for $A^{\rm lens}_{\gamma}$.

Using the expressions above, the matter-matter-lensing bispectrum  reads
\begin{align}
B^{{\rm mml}} ( k_1 , k_2 , k_3 )& =  2 P_{11} ( k_1 ) P_{11 } ( k_2 ) F_2^{\rm lens} ( \kvec_1 , \kvec_2 ) \\ 
& + 2 L_{\rm lens}P_{11} ( k_1 ) P_{11 } ( k_3 ) F_2 ( \kvec_1 , \kvec_3 )   \nonumber  \\   
& + 2 L_{\rm lens} P_{11} ( k_2 ) P_{11 } ( k_3 ) F_2 ( \kvec_2 , \kvec_3 )  \ . \nonumber
\end{align}
\begin{figure*}[t] 
\centering 
\hspace{-.3in} \includegraphics[width=1\textwidth]{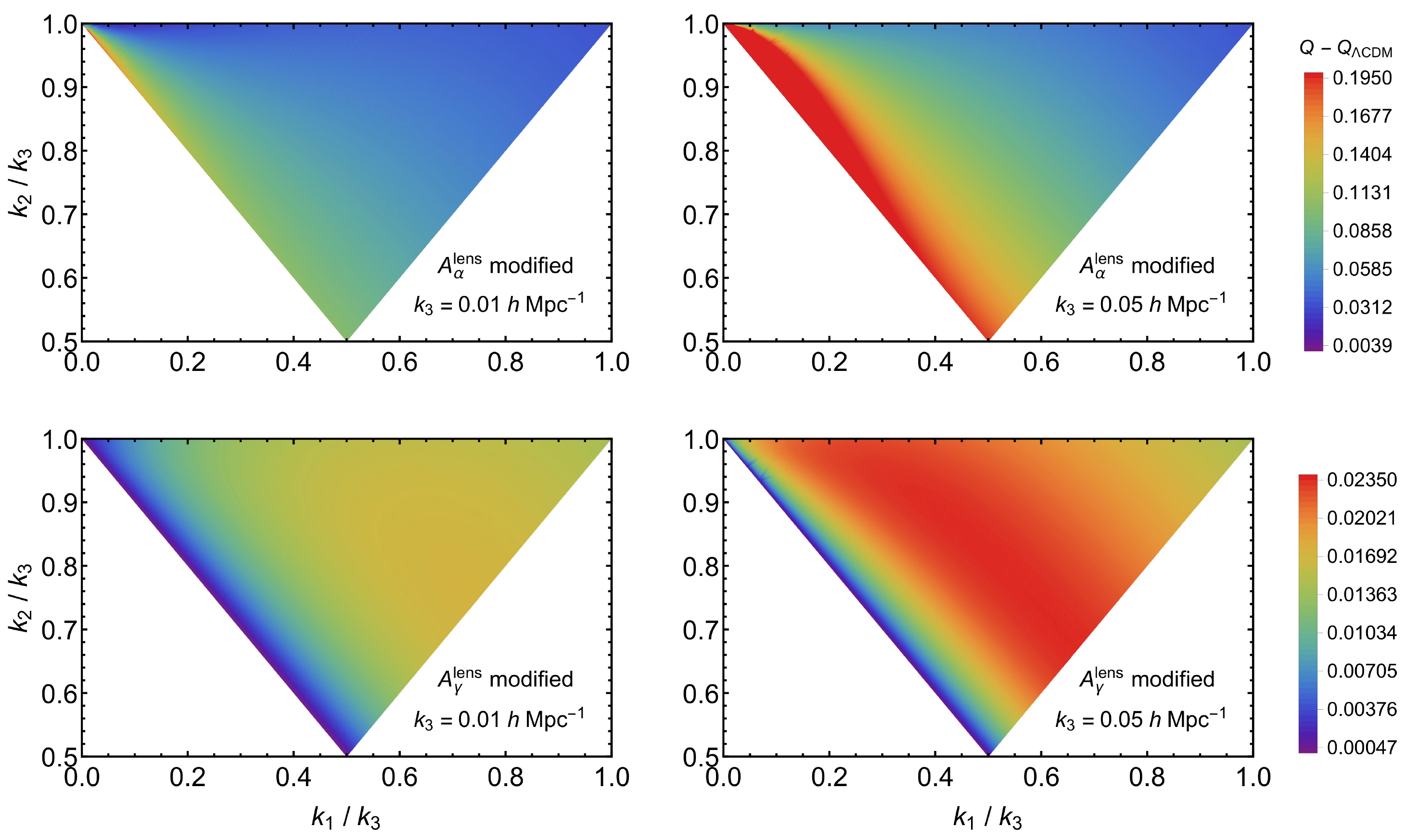} 
\caption{  {Shape of the difference of the reduced cross-correlation bispectrum and the one in $\Lambda$CDM,  $Q^{{\rm mml}} ( k_1 , k_2 , k_3 )- Q^{{\rm mml}}_{\Lambda \text{CDM}} ( k_1 , k_2 , k_3 ) $, for two $10\%$ modifications away from the {$\Lambda$CDM} values  $(A_\alpha^{\rm lens}, A_\gamma^{\rm lens})= (1,-0.284)$, i.e.~for $(A_\alpha^{\rm lens},A_\gamma^{\rm lens}) = (1.1,-0.284)$ (upper panels) and $(A_\alpha^{\rm lens},A_\gamma^{\rm lens})  = (1,-0.256)$  (lower panels), while setting $L_{\rm lens}=1$ and  $(A_\alpha,A_\gamma)$ to their $\Lambda$CDM values.   
Only changing $A_\alpha^{\rm lens}$ produces a signal for folded triangles. Moreover, as discussed in the text,  for $L_{\rm lens}  A_\alpha  - A^{\rm lens}_\alpha \neq 0$ (upper panels) the signal is maximized in the squeezed limit (upper-left corner; since the bispectrum diverges in the squeezed limit,  for presentation purposes we stop plotting it when $Q^{{\rm mml}} - Q^{{\rm mml}}_{\rm \Lambda CDM} > 0.195$). } }
\label{fig:Fig2}
\end{figure*}
{In Fig.~\ref{fig:Fig2} we plot the relative difference between the amplitude of the {reduced} cross-correlation bispectrum, 
\be
Q^{{\rm mml}} ( k_1 , k_2 , k_3 ) \equiv \frac{B^{{\rm mml}} ( k_1 , k_2 , k_3 )}{P_{11} ( k_1) P_{11} ( k_2 ) + ( 2 \text{ perms.})} \;,
\ee
and the one of the {reduced}  cross-correlation  bispectrum in $\Lambda$CDM  as a function of the shape,  for $k_3=0.01 \, h\text{Mpc}^{-1}$ and $k_3=0.05 \, h\text{Mpc}^{-1}$. For simplicity, we set $L_{\rm lens}=1$ and $A_\alpha$ and $A_\gamma$ to their $\Lambda$CDM values, and focus on the effect of  modifications of $A_\alpha^{\rm lens}$ and $A_\gamma^{\rm lens}$.
To show the effect of DHOST theories, in the upper panels we plot this difference in the case where $A_\alpha^{\rm lens}$ is modified by $10\%$ from its $\Lambda$CDM value while $A_\gamma^{\rm lens}$ is unmodified. 
In the lower panels we consider the case where $A_\gamma^{\rm lens}$ is (negatively) modified by $10\%$ from its $\Lambda$CDM value while $A_\alpha^{\rm lens}=1$, as predicted by Horndeski theories. 

As in the case of the auto-correlation bispectrum, only by changing  $A_\alpha^{\rm lens}$ can one  affect the bispectrum for folded triangles. Moreover, contrarily to the auto-correlation case,
changing $A_\alpha^{\rm lens}$
also generates {an enhanced} signal in the squeezed limit.
In particular, in this limit with $\vec q = - \kvec_1$ and $\kvec =  \kvec_2 - \qvec/2$, one has,} up to corrections of order ${\cal O}((q/k)^0)$,
\be \label{squeezemml}
\lim_{q \to 0} \frac{B^{{\rm mml}} (q,k_2,k_3)}{P_{11} ( q ) P_{11} ( k) } \approx  \left(    L_{\rm lens}  A_\alpha  - A^{\rm lens}_\alpha  \right)  \frac{\qvec \cdot \kvec}{q^2} \ . 
 \ee
 The $k/q$ enhancement on the right-hand side shows that the consistency relation does not hold in beyond Horndeski theories, similarly to what happens in the presence  of a violation of the equivalence principle due to a fifth-force \cite{Creminelli:2013nua}. One can check, instead, using the definition of $L_{\rm lens}$,  $L_{\varphi_a}$, and $A_\alpha^{\rm lens}$, respectively eqs.~\eqref{Llens}, \eqref{Ldefs} and \eqref{Alens}, that for Horndeski theories the right-hand side of this equation vanishes, as expected by the consistency relations.  Although, as explained after \eqn{lenspot}, $B^{\rm mml}$ is not directly an observable, \eqn{squeezemml} gives an estimate of what one would find for the bispectrum obtained by cross-correlating observable quantities such as the galaxy-galaxy-galaxy lensing shear bispectrum. The $k/q$ enhancement makes this effect easier to see, so that one can obtain fairly tight observational bounds on the squeezed limit cross-bispectrum without having to go through a full cosmological analysis.

{Additionally, when the consistency relations are broken, different tracers of the dark-matter distribution can in general have different squeezed limits.  This means that when correlating different tracers, one expects an effect of the form \eqn{squeezemml}, proportional to the difference in the bias coefficients of the two tracers.   }

%%%%%%%%%%%%%%
%
\section{Other observational consequences}
\label{sec:observations}

%FV
%The breaking of the consistency relations in DHOST theories has a number of observational consequences on cosmological observables that we briefly discuss in this section.  
The breaking of the consistency relations in DHOST theories has observational consequences on cosmological observables involving higher-order kernels in perturbation theory,  which we briefly discuss in this section. 

\subsection{$n$-point functions}
\label{trispectrum}

Let us discuss the observational consequences associated with squeezed configurations of $n$-point functions.
One can convince oneself that, by symmetry, any {($n+1$)}-point correlation function {of all the same fields} where only one leg  with momentum $q$ is made soft will be proportional to the sum of the short momenta, $\sum_{i=1}^n \kvec_i$, which vanishes at leading order in $q$ by momentum conservation. Thus, as for the (auto-correlation) bispectrum there are no obvious consequences in the single soft-mode squeezed limit of any {($n+1$)}-point correlation function.

As a next possibility, one can consider a higher number of soft modes. 
%FV
The simplest example is  the trispectrum,
 defined by 
\begin{align}
\langle \delta(\kvec_1) \delta(\kvec_2) \delta(\kvec_3) \delta(\kvec_4) \rangle = \ & (2 \pi)^3 \delta_D ( \kvec_1 + \kvec_2 + \kvec_3+ \kvec_4 ) \nonumber \\
& \times T ( \kvec_1 , \kvec_2 , \kvec_3, \kvec_4) \;.
\end{align}
In particular, let us focus on the double-soft limit $T ( \qvec_1,\qvec_2,\kvec_1,\kvec_2)$, where two of the modes, $\qvec_1 $ and $\qvec_2$, are made much smaller than the other two, $\kvec_1$ and $\kvec_2$. 

In the standard case ($\Lambda$CDM or Horndeski theories), the consistency relations ensure that the trispectrum vanishes at leading order in $q_{1,2}/k_{1,2}$ in the double-squeezed limit. 
This is straightforward to verify in perturbation theory. Following the discussion of Sec.~2.1 of \cite{Creminelli:2013poa},   the double-soft limit of the trispectrum in perturbation theory is given by the sum of three contributions.  
%FV
The first is obtained when the density perturbations of the short modes are both taken at second order, i.e., 
\begin{align}
T_{1122} = \ &4 P_{11}(q_1) P_{11}(q_2) P_{11}(| \kvec_1 + \qvec_1| )  F_2(- \qvec_1, \kvec_1 + \qvec_1) \nonumber \\
&\times  F_2(- \qvec_2, \kvec_2 + \qvec_2) +(\kvec_1 \leftrightarrow \kvec_2) \;.
\end{align}   
Another one is obtained when one of the short-mode density perturbation is taken at third order. Defining the third-order kernel $F_3$ analogously to $F_2$ as
\be
\delta^{(3)} ( \kvec) = \int^{\kvec}_{\kvec_1, \kvec_2 , \kvec_3} F_3 ( \kvec_1 , \kvec_2 , \kvec_3) \delta^{(1)}(\kvec_1 ) \delta^{(1)} ( \kvec_2 ) \delta^{(1)}( \kvec_3) \ , 
\ee
(from the definition $F_3$ is symmetric under permutations of $\{ \kvec_1 , \kvec_2  , \kvec_3\}$), this reads
 \be \label{t1113}
T_{1113} = 6  P_{11} ( q_1 ) P_{11}(q_2 ) P_{11} ( k_1 )  F_3 ( -\qvec_1 , -\qvec_2 , -\kvec_1 ) \ .
\ee
The last contribution is obtained by taking the other short mode at third order, {i.e. exchanging $\kvec_1 \leftrightarrow \kvec_2$ in \eqn{t1113}}. 
In the  standard case $F_2(- \qvec_1, \kvec_1 + \qvec_1) \approx - \qvec_1 \cdot \kvec_1/(2 q_1^2) $,   and $F_3(- \qvec_1,- \qvec_2, - \kvec_1 ) \approx (\qvec_1 \cdot \kvec_1)(\qvec_2 \cdot \kvec_1)/({6} q_1^2  q_2^2) $, so that once one considers the permutations, 
there is a cancellation between these three contributions.

This is no longer true in DHOST theories.  If we define the {(time-dependent)} coefficient of  $ F_3$ in the squeezed limit as $B_\alpha $ from $F_3(- \qvec_1,- \qvec_2, - \kvec_1 ) \approx   B_\alpha   (\qvec_1 \cdot \kvec_1)(\qvec_2 \cdot \kvec_1)/({6} q_1^2  q_2^2) $,  for the trispectrum in the double-squeezed limit (setting $\kvec_2 \approx -\kvec_1$) we obtain
\be
\lim_{q_{1,2} \to 0} \frac{T(\qvec_1, \qvec_2, \kvec_1, \kvec_2)}{P(q_1) P(q_2) P(k_1)} \approx  - 8 (A_\alpha^2 - B_\alpha) \frac{ \qvec_1 \cdot \kvec_1}{ 2 q_1^2} \frac{\qvec_2 \cdot \kvec_1}{ 2 q_2^2} \;,
\ee
which shows that the consistency relation is violated in this case. One can show by an explicit computation that  in DHOST theories $B_\alpha \neq A_\alpha^2$ but we postpone its presentation to future work.}

\subsection{Loops}
\label{loops}

The cancellation between $T_{1122}$ and $T_{1113}$ discussed above is also crucial in loop diagrams.\footnote{{For example, the leading IR part of the 1-loop power spectrum can be obtained from the double-soft four-point function by gluing together the two soft legs. }} For instance, the 1-loop power spectrum receives two contributions,
{$P_{1\text{-loop}} = P_{22}+ P_{13}$, where}
\begin{align}
P_{22} (k) &\equiv 2 \int \frac{d^3 \qvec}{(2 \pi)^3} [ F_2(\qvec, \kvec - \qvec) ]^2 P_{11}(q) P_{11} ( |  \kvec - \qvec|)  \;,  \nonumber \\
P_{13} (k)&\equiv  6 \int \frac{d^3 \qvec}{(2 \pi)^3} F_3(\qvec, - \qvec, \kvec ) P_{11}(q) P_{11}(k) \; . 
\end{align} 
In the standard case, the IR parts of these integrals, coming from the  small momenta 
%FV
$q \ll k$, 
cancel when summing $P_{22}+ P_{13}$ 
%FV
\cite{Jain:1995kx,Scoccimarro:1995if,Carrasco:2013sva,Blas:2013aba} as a consequence of the equivalence principle \cite{Creminelli:2013mca}. 
This was also shown to happen in $\Lambda$CDM and quintessence theories with exact time dependence \cite{Lewandowski:2017kes}.

This cancellation does not hold anymore when the consistency relations are violated \cite{Creminelli:2013poa}. Indeed, as expected from the above discussion, in DHOST theories we have 
%FV
\be
\frac{ P_{1\text{-loop}} (k)}{ P_{11}(k) }   \approx 4 (A_\alpha^2 -B_\alpha) \int_{{q \lesssim k}} \frac{d^3 \qvec}{(2 \pi)^3} \bigg( \frac{\qvec \cdot \kvec}{2 q^2} \bigg)^2 P_{11}(q)  \;.
\ee
Here the IR divergences come from the $\qvec \rightarrow 0$ limit in $P_{13}$, and from the $\qvec \rightarrow 0$ and $\qvec \rightarrow \kvec$ limits in $P_{22}$. This expression can be rewritten as
\be
\begin{split}
\frac{ P_{1\text{-loop}} (k)}{ P_{11}(k) }  \approx (A_\alpha^2 -B_\alpha)  \frac{k^2}{3}\sigma^2_v\;, 
\end{split}
\ee
where $\sigma^2_v \equiv \int_{{q \lesssim k}} \frac{d^3 \qvec}{(2 \pi)^3} P_{11}(q) /q^2 $ is the variance of the long-wavelength displacement, which depends on where we place the IR cutoff. Taking it for instance at $0.04 \, \text{Mpc}^{-1} h$ we find $\sigma_v^{-1} \simeq 0.15\, \text{Mpc}^{-1} h$, which is where non-linear scales start. {(For power-law universes with $P_{11} ( k ) \propto k^n$, this loop will only be IR-convergent for $n>-1$ instead of the standard $n>-3$ for theories that satisfy the equivalence principle.)}

In conclusion, in $\Lambda$CDM and Horndeski theories (i.e.~for $A_\alpha = B_\alpha =1$) the long-wavelength displacement of momentum $q$ does not affect the 1-loop power spectrum at $k \gg q$, as  expected from the equivalence principle, but in DHOST theories 
the  long-wavelength motion  affects the short-scale physics on non-linear scales.

\section{Conclusion}
\label{sec:conclusion}

In this paper we have investigated the consistency relations for large-scale structure when gravity is modified. 
Assuming Gaussian initial conditions, we have shown that the consistency relations derived in $\Lambda$CDM also hold for Horndeski theories. Indeed, the gravitational and fluid  equations are simultaneously invariant under the same combination of coordinate and field transformations, which we discuss in Secs.~\ref{leadingirtermssec} and \ref{symmIR}, that can be used to remove 
the effect of a long-wavelength mode $q$ on the short-scale physics (see Sec.~\ref{CRHorndeski}). This is analogous to what happens in $\Lambda$CDM due to the equivalence principle.

The validity of the consistency relations {for Horndeski theories} was derived using perturbation theory, but as long as the coupling with the long wavelength mode satisfies the equivalence principle, these relations are non-perturbative, i.e.~they hold regardless of the non-linear short-scale physics, such as baryonic effects, bias, etc. 
Of course, in scalar-tensor theories, self-interactions of the scalar field are known to renormalize the coupling to a long-wavelength field and lead to violations of the equivalence principle \cite{Hui:2009kc}. However, the class of theories that we discuss here enjoy Galilean invariance \cite{Nicolis:2008in} and are known not to renormalize the scalar charge as long as the gravitational and scalar binding energies are negligible \cite{Hui:2009kc,Hui:2010dn}.

Next, we extended this study to DHOST theories. As discussed in Sec.~\ref{dhostsymm}, in this case, due to the kinetic coupling between matter and the scalar field, the gravitational and fluid  equations are  invariant under two separate combinations of coordinate and field transformations. In the absence of a common  transformation, the consistency relations are not satisfied. 

We have also discussed the consequences  on the perturbation-theory kernels. In $\Lambda$CDM, their values in the squeezed limit (i.e.~in the limit when one or more modes become much smaller than the others) are protected by the symmetry transformation of the gravitational and fluid  equations. We have shown that these properties also extend to Horndeski theories. However, in theories beyond Horndeski the perturbation-theory kernels   are modified also in the squeezed limit, as we have shown explicitly by computing the second-order solutions of the density contrast and velocity divergence.
Although  { a large-scale physical adiabatic mode} is absent in this case, we can still use the transformations of the gravitational field and fluid equations  to compute the form of the second-order kernels in the squeezed limit from the linear solution, see Sec.~\ref{symmIR}. 
%The leading non-linear IR terms continue to be fixed by the linear ones.

Using the second-order kernel, in Sec.~\ref{bispectrumsec1} we have computed the matter bispectrum. Thanks to translational invariance,  it is not enhanced as $1/q$ in the squeezed limit even in beyond Horndeski theories. However, for DHOST theories, its shape receives modifications in the so-called folded limit $k_1+k_2=k_3$, confirming and extending the results of \cite{Hirano:2018uar}.
To see an enhancement signalling a violation of the consistency relation, we must consider the cross-correlation between different tracers, such as for instance the correlation of the density contrast with the lensing potential, computed in Sec.~\ref{sec:lensing}. 

Violation of the consistency relations can be  also observed in higher-order correlators and in  Sec.~\ref{trispectrum} we study these violations in the trispectrum in the double squeezed limit. The same effects can be also observed in the 1-loop power spectrum (Sec.~\ref{loops}): {while} in $\Lambda$CDM and in Horndeski theories short-scale physics is not affected by longer modes as expected by the equivalence principle, in DHOST theories this is not the case. 
For this reason, the usual treatment of BAO reconstruction and IR resummation (see for instance \cite{Lewandowski:2018ywf} and references therein for past and recent developments) cannot be applied to these theories and must be revised. We postpone a detailed study of these topics to future work.

\section*{Acknowledgements}
We thank Emilio Bellini, Austin Joyce, Marcel Schmittfull, Leonardo Senatore, Zack Slepian and Miguel Zumalacarregui for interesting correspondence and discussions.  We particularly 
thank Lam Hui and Marko {Simonovi\'c}  for insightful discussions and comments on the article. M.~C.~is supported by the Labex P2IO. M.~C.~thanks SISSA for the warm hospitality during the last stage of this work.  M.~L.~acknowledges financial support from the European Research Council under ERC-STG-639729, \emph{preQFT: Strategic Predictions for Quantum Field Theories}.  F.V.~acknowledges partial support by the Munich Institute for Astro- and Particle Physics (MIAPP) of the DFG cluster of excellence ``Origin and Structure of the Universe,'' where this work was completed.

%%%%%%%%%%%%%%%%%%%%
%
%
%
%
%
%
\appendix

\section{Coefficients} 
\label{sec:coeff}
Here, we report the { covariant DHOST action, its expression in terms of the EFT of DE, and the} explicit expressions for many of the coefficients appearing in the main text.  We find it useful sometimes to write $\chi_N = \{ \Phi , \Psi \}$ in order to separate the Newtonian potentials from the scalar field $\pi$.  

{
\subsection{DHOST action and map} 
\label{dhostactionmap}
The action for DHOST theories, including all possible quadratic combinations up to second derivatives of the field $\phi$, reads
\begin{align}
\label{TypeIa}
& S_{\rm DHOST} = \ \int d^4 x \sqrt{-g} \Big[ P(\phi,X)+ Q(\phi,X) \Box \phi \\ 
& \quad \quad \quad \quad + f(\phi,X) {}^{(4)}\!R + \sum_{I=1}^5 a_I (\phi,  X ) L_I (\phi, \phi_{;\nu},  \phi_{;\rho \sigma})  \Big] \;, \nonumber
\end{align}
where $X \equiv - \phi_{;\mu} \phi^{;\mu}/2 $, {a semicolon denotes the covariant derivative}, ${}^{(4)}\!R$ is the 4D Ricci scalar, $P, Q, f$ and $a_I$ are free functions, 
 and the $L_I$ are defined by
\begin{align}
L_1&= \phi_{;\mu\nu}\phi^{;\mu\nu} \;, \quad  L_2=(\phi^{;\mu}_{;\mu})^2 \;, \quad  L_3=(\phi^{;\mu}_{;\mu})(\phi^{;\rho} \phi_{;\rho\sigma}\phi^{;\sigma})\;, \nonumber \\
  L_4&= \phi^{;\mu}\phi_{;\mu\nu}\phi^{;\nu\rho}\phi_{;\rho} \;, \quad  L_5=(\phi^{;\rho} \phi_{;\rho\sigma}\phi^{;\sigma})^2 \;.
\end{align}

The time-dependent functions appearing in the EFT of DE action \eqn{EFTaction} are related to the free functions in the above action as \cite{Dima:2017pwp, Crisostomi:2019yfo}
\begin{align}
\label{alphabeta}
M^2 & = 2( f - 2 a_2 X) \;,  \nonumber \\
\alphaB & = \alphaV -3 \beta_1 + \dot \phi ( f_\phi + 2 X f_{,\phi X} + X Q_{,X}) /(M^2H) \;, \nonumber  \\
c_T^2 &= {2 f}/{M^2}\,, \nonumber  \\
  \alphaH & = 4 X (a_2 - f_{,X})/M^2  \,,\nonumber \\
\bun& =  2 {X} (f_{,X}- a_2+ a_3 X)/{M^2} \,, \nonumber \\
\bdeux & =  -  {8X^2} \left(a_3+a_4  - 2 a_5X\right)/{M^2} \,, \nonumber \\
\btrois &=-  {8X}( f_{,X} -  a_2-  a_4 X)/{M^2} \, ,  \nonumber   \\
\alphaV & = {4 X ( f_{,X} - 2a_2 -2X a_{2,X} )}/{M^2} \;.
\end{align}
}

%%%%%%%%%%%%%%%
%
%

\subsection{Equations of motion} 
\label{coefficientsapp}
The coefficients appearing in the equations of motion \eqns{phieq1}{pieq1}, are given explicitly by 
 \begin{align}
c_1 & = -4H \alphaB  + H ( 4 \alphaH - 2 \beta_3 (1 + \alphaM )) - 2 \dot \beta_3 \;,   \\
c_2 & = 4 H (1 + \alphaM - c_{\rm T}^2)  +  4 \left(  H    \alphaH (1 + \alphaM)   + \dot \alpha_{\rm H} \right)\;,  \nonumber\\
c_3 & = - 2 H^2 \mathcal{C}_2+ \frac{1}{2} \left\{   H \left[ 4 \dot \alpha_{\rm H} - 2 ( 1 + \alphaM) \dot \beta_3  - \beta_3 \dot {\alpha}_{\rm M} \right] - \ddot \beta_3           \right\}   \nonumber  \\ 
& \quad \quad + \frac{1}{2} \big\{ - H^2 ( 1 + \alphaM) \left[ - 4 \alphaH + \beta_3 ( 1 + \alphaM)\right]   \nonumber \\
& \quad \quad  + 4 \alphaH \dot H - \beta_3 ( 1 + \alphaM) \dot H  \big\} \;,  \nonumber \\
 c_4 & = 4 ( 1 + \alpha_{\rm H} ) \ ,  \qquad  c_5  = -2 c_{\rm T}^2 \ , \qquad  c_6  = - \beta_3 \ ,  \nonumber\\ 
 c_7  &= 4 \alpha_{\rm H} \;, \qquad
 c_8  = -2 (  2 \beta_1 + \beta_3 ) \ , \qquad  c_9  = 4 \beta_1 + \beta_3 \ , \nonumber
 \end{align}
 with
\begin{align}
\begin{split}
\mathcal{C}_2 & \equiv  - \alphaM + \alphaB ( 1 + \alphaM ) + c_{\rm T}^2 -1   \\
& \quad \quad +  ( 1 + \alphaB)\frac{\dot H}{H^2} + \frac{ \dot \alpha_{\rm B} }{H} + \frac{\bar \rho_{\rm m}}{2H^2 M^2}   \ ,
\end{split}
\end{align}
 \be
 \begin{split}
 b_1 & =  H  \left[ 4 \alphaB + \alphaV (-1+\alphaM)  -2 \alphaM + 3 (c_{\rm T}^2 -1) \right] \\
 & \quad  + \dot \alpha_{\rm V}   - H \left[  8 \beta_1 \alphaM + \alphaH ( 3 + \alphaM ) \right]  - \dot \alpha_{\rm H} - 8 \dot \beta_1 \;, \\
 b_2 & = \alpha_{\rm V} - \alpha_{\rm H} - 4 \beta_1 \ , \qquad  b_3  = c_{\rm T}^2 -1  \ .
 \end{split}
\ee
We have also defined 
\begin{align}
C_1 & \equiv  \frac{1}{4} \left( c_1  - H c_8 ( 1 + \alphaM)  - \dot c_8 \right) \;,   \nonumber \\
C_2 & \equiv \frac{1}{4} \left(   c_2 - H c_7 ( 1 + \alphaM ) - \dot c_7 \right)\;, \nonumber \\
C_3 & \equiv  \frac{1}{4} \Big\{ 2 c_3 + (1 + \alphaM) \left[ 2 H \dot c_9 + c_9 \left( H^2 (1 + \alphaM)  + \dot H \right)  \right]  \nonumber \\
& \quad \quad \quad  + c_9 H \dot \alpha_{\rm M} + \ddot c_9     \Big\}  \;, \nonumber \\
C_4 & \equiv \half ( c_9 H ( 1 + \alphaM) + \dot c_9 ) \;.
\end{align}
Note that these equations are general: no degeneracy conditions or observational constraints have been assumed.

%%%%%%%%%%%%%
%
%
\subsection{Linear solutions} \label{linearsolssec}

Here, we focus on the linear equations of motion to give explicit expressions for the coefficients appearing in \eqn{linearsols}.  As usual, we solve \eqns{phieq1}{psieq1} for $\partial^2 \Phi$ and $\partial^2 \Psi$ in terms of $\partial^2 \pi$, $\partial^2 \dot \pi$, and $\delta$.  This has the form
\begin{align}
\begin{split}
\partial^2 \chi_N = \omega^{ \chi_N}_{1} \partial^2 \pi + \omega^{ \chi_N}_2 \partial^2 \dot \pi + \omega^{ \chi_N}_3 \delta  \ , 
\end{split}
\end{align}
where 
\begin{align}
\begin{split}
& \omega^\Phi_1 = \frac{ 8 C_1 c_5 -4 C_2 c_4 }{\omega} \ , \quad \omega^\Phi_2 = \frac{c_4 c_7 - 2 c_5 c_8 }{\omega} \\
& \omega^\Phi_3 = \frac{-4 a^2 c_5 \barrhom}{\omega M^2}  \ , \quad \omega^\Psi_1 = \frac{ 8 C_2 c_6  -4 C_1 c_4 }{\omega}  \\
& \omega^\Psi_2 = \frac{c_4 c_8 - 2 c_6 c_7 }{\omega} \ , \quad \omega^\Psi_3 = \frac{2 a^2 c_4 \barrhom }{\omega M^2} \ ,
\end{split}
\end{align}
where $\omega \equiv c_4^2 - 4 c_5 c_6$.  
Next, we plug these expressions into \eqn{pieq1} to obtain the expression for $\partial^2 \pi$, where, once we impose the degeneracy conditions discussed in \secref{eomsec}, the terms proportional to $\partial^2 \dot \pi$ and $\partial^2 \ddot \pi$ drop out, and we are left with an expression as in \eqn{linearsols} with
\begin{align}
\begin{split}
\mu_\pi &  =-  \frac{c_1\omega^\Phi_3 + c_2 \omega^\Psi_3 + c_8 \dot \omega^\Phi_3 + c_7 \dot \omega^\Psi_3}{ a^2 C_\pi} \ , \\
\nu_\pi  & = -  \frac{c_8 \omega^\Phi_3 + c_7 \omega^\Psi_3 }{a^2 C_\pi} \ , 
\end{split}
\end{align}
where we have defined
\be
C_\pi = 4 C_3 + c_1 \omega^\Phi_1 + c_2 \omega^\Psi_1 + c_8 \dot \omega^\Phi_1 + c_7 \dot \omega^\Psi_1 \ . 
\ee

 Next, we plug the solution for $\partial^2 \pi$ into \eqns{phieq1}{psieq1} to get the solutions for $\partial^2 \Phi$ and $\partial^2 \Psi$ in the form \eqn{linearsols} with
 \begin{align}
 \begin{split}
 \mu_{\chi_N} &=  \mu_\pi (\omega^{\chi_N}_1 + 2 H \omega^{\chi_N}_2 ) + \dot \mu_\pi  \omega^{\chi_N}_2+ a^{-2}\omega^{\chi_N}_3  \ , \\
 \nu_{\chi_N} &= \mu_\pi \omega^{\chi_N}_2 + \nu_\pi (\omega^{\chi_N}_1 + 2 H \omega^{\chi_N}_2 ) + \dot \nu_\pi \omega^{\chi_N}_2 \ , \\
 \sigma_{\chi_N} & = \nu_\pi \omega^{\chi_N}_2 \ . 
 \end{split}
 \end{align}
Note that these equations are general: no degeneracy conditions or observational constraints have been assumed, except to say that the terms proportional to $\partial^2 \dot \pi$ and $\partial^2 \ddot \pi$ drop out of the solution for $\partial^2 \pi$.  

Now, we specialize to the case where we impose the degeneracy conditions in {\secref{eomsec}}, along with $\alphaH = - 2 \beta_1$ (which imposes that gravitational waves do not decay).\footnote{Notice that, while we use many of the same symbols as \cite{Crisostomi:2019yfo}, their definitions have changed slightly in order to simplify the current work.}  This gives
\begin{align}
\mu_\pi & =  \frac{1}{M^2 C_\pi (1 - \beta_1)^2} \Big(  2(1- \beta_1)\beta_1 \dot{\bar \rho}_{\rm m} \nonumber \\
& \quad \quad + 2 H \barrhom  (\alphaB - \alphaM ( 1 - \beta_1) + \beta_1 (4 - 3 \beta_1))    \Big)  \nonumber \\
\nu_\pi & = \frac{2 \beta_1 \barrhom}{M^2 C_\pi ( 1 - \beta_1 ) } \
\end{align}
where
\be
C_\pi = \frac{2 H^2 \alpha c_s^2}{(1-\beta_1)^2 } \ , 
\ee
and
\begin{align}
& \alpha  c_s^2   \equiv - \frac{\bar \rho_{\rm m} ( 1 - \beta_1)^2}{ H^2 M^2}\\
 & +  {2}  \left( 1 + \alphaB - \frac{\dot \beta_1 }{H} \right)^{2}   \bigg[  \frac{1}{a M^2} \frac{d}{dt }   \left( \frac{a M^2 (1-\beta_1)}{H ( 1 + \alphaB) - \dot \beta_1}  \right)  -1  \bigg]  \nonumber  \ . 
\end{align}

For the other coefficients in \eqn{linearsols}, we have 
\begin{align}
\begin{split}
\mu_{\Phi} &  = \frac{\barrhom}{2 M^2 (1-\beta_1)^2} + \frac{\mu_\pi \varpi_\Phi - \dot \mu_\pi \beta_1 }{1 - \beta_1 } \ , \\
\nu_\Phi & = \frac{ - ( \mu_\pi + \dot \nu_ \pi )\beta_1 + \nu_\pi \varpi_\Phi}{1-\beta_1} \ ,  \,\,\,\, \sigma_\Phi  = - \frac{\nu_\pi \beta_1}{1-\beta_1} \ , 
\end{split}
\end{align}
and 
\begin{align}
\begin{split}
\mu_{\Psi} &  = \frac{\barrhom(1-2 \beta_1)}{2 M^2 (1-\beta_1)^2} + \frac{\mu_\pi \varpi_\Psi + \dot \mu_\pi \beta_1 }{1 - \beta_1 } \ , \\
\nu_\Psi & = \frac{  ( \mu_\pi + \dot \nu_ \pi )\beta_1 + \nu_\pi \varpi_\Psi}{1-\beta_1} \ ,  \quad \sigma_\Psi  =  \frac{\nu_\pi \beta_1}{1-\beta_1} \ , 
\end{split}
\end{align}
where we have defined 
\begin{align}
\varpi_\Phi & = \frac{H(\alphaB - \alphaM -\beta_1 (1 -\alphaM  - 2 \beta_1)) - \dot \beta_1 }{1-\beta_1} \ , \nonumber \\
\varpi_\Psi & = \frac{1}{1-\beta_1} \Big(  - \dot \beta_1 + 2 \beta_1 \dot \beta_1  \\
& + H(\alphaB +\beta_1 (3-2\alphaB +\alphaM ) - \beta_1^2(4 + \alphaM) ) \Big) \ . \nonumber
\end{align}

%%%%%%%%%%%%%%%
%
%
\subsection{Quadratic solutions} \label{quadraticapp}
The solutions for the gravitational potentials in terms of $\pi$ can be written
\begin{align} \label{phipsisol2}
\partial^2 \chi_N^{(2)} &  =  \omega^{\chi_N}_{1} \partial^2 \pi^{(2)} + \omega^{\chi_N}_2 \partial^2 \dot \pi^{(2)} { + \omega_3^{\chi_N} \delta^{(2)}} \\
& + a^{-2} \left( \lambda_1^{\chi_N} Q_2[\pi^{(1)} , \pi^{(1)}] + \lambda^{\chi_N}_2 P_2[\pi^{(1)} , \pi^{(1)} ]   \right)\nonumber \ , 
\end{align}
where we have defined $P_2[\varphi_a , \varphi_b ] =\partial_i \left( \partial_j \varphi_a \partial_i \partial_j \varphi_b \right)$, and
\begin{align}
& \lambda^\Phi_1 = \frac{ 2 b_2 c_5 - b_3 c_4 }{\omega} \ , \quad \lambda^\Phi_2 = \frac{ 2 c_5c_8- c_4 c_7 }{\omega} \\
&  \lambda^\Psi_1 = \frac{  2 b_3 c_6 - b_2 c_4}{\omega} \ , \quad \lambda^\Psi_2 = \frac{  2 c_6 c_7 - c_4 c_8} {\omega} \ .  \nonumber
\end{align}
Now, taking \eqn{phipsisol2}, plugging it into \eqn{pieq1}, and replacing all of the linear solutions in the form $\partial^2 \varphi_a^{(1)} = a^2 L_{\varphi_a} \delta^{(1)}$ using \eqn{Ldefs}, we obtain a solution for $\partial^2 \pi^{(2)}$ of the form \eqn{secondorderform} with
\be \label{upsilonpi}
 \upsilon_\alpha^{\pi} = - \nu_\pi L_\pi \  ,  \quad \text{and} \quad  \upsilon_\gamma^{\pi} = -  \upsilon_\alpha^{\pi} + \Delta \upsilon_\gamma^\pi \ , 
\ee
where
\begin{align}
- \frac{C_\pi}{L_\pi} \Delta \upsilon_\gamma^\pi & =  b_1 L_\pi +2 b_2 L_\Phi + 2 b_3 L_\Psi  \\
&+ q_{1,8,\Phi} + q_{2 , 7 , \Psi}  - 2c_9 ( (2 + f) H L_\pi + \dot L_\pi ) \ , \nonumber
\end{align}
with
\begin{align}
q_{i,j,\chi_N } & = -c_j L_{\chi_N} + L_\pi \lambda_1^{\chi_N} c_i  \\
& + L_\pi \lambda_1^{\chi_N} c_j \left( 2 (1 +f) H   + 2 \frac{ \dot L_\pi}{L_\pi} + \frac{ \dot \lambda_1^{\chi_N} }{\lambda_1^{\chi_N}} \right)    \ . \nonumber
\end{align}

Now, we plug the solution for $\partial^2 \pi^{(2)}$ into \eqn{phipsisol2} to get, 
\begin{align}
\begin{split} \label{upsilonphi}
\upsilon_\alpha^{\chi_N} & = -    \nu_{\chi_N} L_\pi - \sigma_{\chi_N} ( 3 H  f L_\pi + \dot L_\pi    )       \  , \\
\upsilon_\gamma^{\chi_N} & = L_\pi^2 ( \lambda_1^{\chi_N} - \lambda_2^{\chi_N} )  \\
& + \upsilon^\pi_\gamma ( \omega_1^{\chi_N} + 2 (1 + f ) H \omega_2^{\chi_N}  ) + \dot \upsilon_\gamma^\pi \omega_2^{\chi_N}  \ . 
\end{split}
\end{align} 
Note that these equations are general: no degeneracy conditions or observational constraints have been assumed.  Also, notice that the coefficients of $X_{\alpha_s}$ are relatively simple and only depend on the coefficients in the linear equations of motion.  The reason for this is discussed in \secref{leadingirtermssec}.

\section{Green's function}
\label{sec:Green}

\subsection{Definition}
\label{app:green1}
To solve the system \eqn{eom1} perturbatively, we use the Green's function, defined by 
\begin{align}
\begin{split}
\partial_t^2 G ( t , t_1 ) + \bar \nu_\Phi ( t ) \partial_t G( t , t_1) - & \bar \mu_\Phi ( t ) G( t , t_1 )\\
&\quad \quad   = \delta_D ( t - t_1) \ , 
\end{split}
\end{align}
whose explicit expression is given by 
\be
G(t, t_1 ) = \bar G ( t , t_1 ) \Theta_{\rm H} ( t - t_1)  \ ,
\ee
with
\be \label{gfdef}
\bar G( t , t_1 ) \equiv \frac{ D_- ( t ) D_+ ( t_1) - D_+ ( t ) D_- ( t_1) }{W(t_1)} \ , 
\ee
where the Wronskian is given by 
\be
W( t ) = D_+ (t ) \dot D_-(t) - D_-(t) \dot D_+(t) \ ,
\ee
and $\Theta_{\rm H}$ is the Heaviside step function.

\subsection{{Green's function manipulations}}
\label{app:green2}
To solve \eqn{delta2ireq} for $\delta^{(2)}$ in the squeezed limit, we apply the Green's function to the right-hand side.  We first apply it to the second line in \eqn{delta2ireq}, where we will find that we obtain the standard contribution.  We have
\be \label{agfintegral}
\int_0^t d t_1 \bar G ( t , t_1 ) \left( 2 \frac{\partial_i \dot \delta }{\partial^2} \partial_i \dot \delta  + \frac{\partial_i \ddot \delta}{\partial^2 } \partial_i \delta  + \bar \nu_\Phi  \frac{\partial_i \dot \delta}{ \partial^2 } \partial_i \delta   \right)_{t_1} \ .
\ee
Next, we integrate by parts the term proportional to $\ddot \delta$ to get 
\begin{align} \label{nextstep}
&\int_0^t d t_1 \, \frac{ \partial_i \dot \delta ( t_1)}{\partial^2} \Big\{ \left( \bar \nu_\Phi ( t_1) \bar G( t , t_1) - \partial_{t_1} \bar G(t , t_1) \right) \partial_i \delta ( t_1)  \nonumber \\
& \hspace{1.5in} + \bar G ( t , t_1) \partial_i \dot \delta ( t_1 )     \Big\} \ ,
\end{align}
{and we remind the reader that all $\delta$ fields appearing above are the linear field $\delta^{(1)}$.  }

Now, one can check explicitly using \eqn{gfdef} and the fact that $\dot W = - \bar \nu_\Phi W$, that 
\begin{align}
\begin{split} \label{gfprop1}
& \bar \nu_\Phi ( \ttt) \bar G( t , \ttt)- \partial_{\ttt} \bar G( t , \ttt) = \\
 & \quad \quad \quad W(\ttt)^{-1} \left( D_+ ( t ) \dot D_- ( \ttt) - D_- ( t ) \dot D_+ ( \ttt) \right) 
 \end{split}
\end{align}
and further that 
\begin{align} \label{gfprop2}
& W(\ttt)^{-1} \left( D_+ ( t ) \dot D_- ( \ttt) - D_- ( t ) \dot D_+ ( \ttt) \right)  \delta^{(1)} ( t_1 ) = \nonumber \\
& \hspace{1.3in}  \delta^{(1)} ( t )  - \bar G( t  , t_1 ) \dot \delta^{(1)}( t_1) \ ,
\end{align}
for any linear combination of growing and decaying modes $\delta^{(1)}$.  
Using \eqn{gfprop1} and \eqn{gfprop2}, \eqn{nextstep} becomes
\be \label{lcdmdipole}
\int_0^t d t_1 \frac{ \partial_i \dot \delta^{(1)} ( t_1)}{\partial^2}  \partial_i \delta^{(1)} ( t ) =  \frac{ \partial_i  \delta^{(1)} ( t)}{\partial^2} \partial_i   \delta^{(1)} ( t )  \ ,
\ee
which is the standard contribution familiar from $\Lambda$CDM and Horndeski theories.

Finally, we apply the Green's function to the last two lines of \eqn{delta2ireq}.  This does not simplify in any particularly nice way, so, after plugging in the linear fields, we get \eqn{delta2irsol}.

{With these manipulations in mind, we can also show how to directly obtain the identity \eqn{aalphaid}.  For this, we use the growing mode solution \eqn{lineardelta} in \eqn{agfintegral}, from which we find
\begin{align}
1 &= \int_0^t d t_1 \bar G( t , t_1 ) \frac{D_+(t_1)^2}{D_+ (t)^2} \\
& \quad \quad \quad   \times \left( 3 f^2 H^2 + H \dot f + f \dot H + \bar \nu_\Phi H f  \right)_{t_1} \ . \nonumber
\end{align}
Now, we would like to rewrite \eqn{upsilondeltaalpha} for $\upsilon_\alpha^\delta$ so that we isolate the expression above in parentheses, so we add and subtract $3 f^2 H^2 + H \dot f + f \dot H + \bar \nu_\Phi H f $ to  the right-hand side of \eqn{upsilondeltaalpha} to get 
\begin{align}
\upsilon_\alpha^\delta = 3 f^2 H^2 + H \dot f + f \dot H + \bar \nu_\Phi H f  + K_2 \ .
\end{align}
After plugging this into \eqn{a1and2}, we obtain \eqn{aalphaid}.  
}

\newpage

\bibliographystyle{utphys}
\bibliography{EFT_DE_biblio3}

\end{document}